\documentclass[twocolumn,twocolappendix]{aastex63}

\received{Month Day, 2020}
\revised{Month Day, Year}
\accepted{Month Day, Year}
\submitjournal{ApJS}

\newcommand{\zYJ}{{z^{\prime}YJ}}
\newcommand{\loggf}{\log gf}

\newcommand{\FeH}{{\rm [Fe/H]}}
\newcommand{\XH}[1]{{\rm [{#1}/H]}}
\newcommand{\XFe}[1]{{\rm [{#1}/Fe]}}
\newcommand{\Teff}{T_{\rm eff}}
\newcommand{\epsX}[1]{\epsilon({\rm {#1}})}
\newcommand{\EP}{{\rm EP}}
\newcommand{\Lc}{\lambda_{\rm c}}
\newcommand{\fsyn}{f_{\rm syn}}
\newcommand{\fsynr}{f_{\rm syn}^\dagger}

\newcommand{\kms}{{\rm km~s^{-1}}}

\newcommand{\Dtot}{{\Delta_{\rm tot}}}
\newcommand{\Dmed}{{\Delta_{\rm med}}}

\newcommand{\DT}{{\Delta_{\Teff}}}

\newcommand{\DG}{{\Delta_{\log g}}}

\newcommand{\DZ}{{\Delta_{\XH{M}}}}

\newcommand{\DX}{{\Delta_{\xi}}}

\defcitealias{Kondo-2019}{Paper~I}

\shorttitle{Near-infrared lines of six elements in addition to iron}
\shortauthors{Fukue et al.}
\begin{document}

\title{Absorption Lines in the 0.91--1.33\,{$\mu$}m Spectra of Red Giants for Measuring \\ Abundances of Mg, Si, Ca, Ti, Cr, and Ni}

\correspondingauthor{Kei Fukue, Noriyuki Matsunaga}
\email{k.fukue@cc.kyoto-su.ac.jp, matsunaga@astron.s.u-tokyo.ac.jp}

\author{Kei Fukue}
\affiliation{Laboratory of Infrared High-resolution spectroscopy (LiH), Koyama Astronomical Observatory, \\ Kyoto Sangyo University, Motoyama, Kamigamo, Kita-ku, Kyoto 603-8555, Japan}

\author{Noriyuki Matsunaga}
\affiliation{Department of Astronomy, School of Science, The University of Tokyo,
7-3-1 Hongo, Bunkyo-ku, Tokyo 113-0033, Japan}
\affiliation{Laboratory of Infrared High-resolution spectroscopy (LiH), Koyama Astronomical Observatory, \\ Kyoto Sangyo University, Motoyama, Kamigamo, Kita-ku, Kyoto 603-8555, Japan}

\author{Sohei Kondo}
\affiliation{Kiso Observatory, Institute of Astronomy, School of Science, The University of Tokyo, 10762-30 Mitake, Kiso-machi, Kiso-gun, Nagano 397-0101, Japan}
\affiliation{Laboratory of Infrared High-resolution spectroscopy (LiH), Koyama Astronomical Observatory, \\ Kyoto Sangyo University, Motoyama, Kamigamo, Kita-ku, Kyoto 603-8555, Japan}

\author[0000-0002-2861-4069]{Daisuke Taniguchi}
\affiliation{Department of Astronomy, School of Science, The University of Tokyo,
7-3-1 Hongo, Bunkyo-ku, Tokyo 113-0033, Japan}

\author[0000-0003-2380-8582]{Yuji Ikeda}
\affiliation{Photocoding, 460-102 Iwakura-Nakamachi, Sakyo-ku, Kyoto 606-0025, Japan}
\affiliation{Laboratory of Infrared High-resolution spectroscopy (LiH), Koyama Astronomical Observatory, \\ Kyoto Sangyo University, Motoyama, Kamigamo, Kita-ku, Kyoto 603-8555, Japan}

\author[0000-0003-4578-2619]{Naoto Kobayashi}
\affiliation{Kiso Observatory, Institute of Astronomy, School of Science, The University of Tokyo, 10762-30 Mitake, Kiso-machi, Kiso-gun, Nagano 397-0101, Japan}
\affiliation{Institute of Astronomy, School of Science, The University of Tokyo, 2-21-1 Osawa, Mitaka, Tokyo 181-0015, Japan}
\affiliation{Laboratory of Infrared High-resolution spectroscopy (LiH), Koyama Astronomical Observatory, \\ Kyoto Sangyo University, Motoyama, Kamigamo, Kita-ku, Kyoto 603-8555, Japan}

\author[0000-0001-6401-723X]{Hiroaki Sameshima}
\affiliation{Institute of Astronomy, School of Science, The University of Tokyo, 2-21-1 Osawa, Mitaka, Tokyo 181-0015, Japan}

\author[0000-0002-6505-3395]{Satoshi Hamano}
\affiliation{National Astronomical Observatory of Japan, 2-21-1 Osawa, Mitaka, Tokyo 181-8588, Japan}

\author{Akira Arai}
\affiliation{Laboratory of Infrared High-resolution spectroscopy (LiH), Koyama Astronomical Observatory, \\ Kyoto Sangyo University, Motoyama, Kamigamo, Kita-ku, Kyoto 603-8555, Japan}

\author{Hideyo Kawakita}
\affiliation{Laboratory of Infrared High-resolution spectroscopy (LiH), Koyama Astronomical Observatory, \\ Kyoto Sangyo University, Motoyama, Kamigamo, Kita-ku, Kyoto 603-8555, Japan}
\affiliation{Department of Physics, Faculty of Science, Kyoto Sangyo University, Motoyama, Kamigamo, Kita-ku, Kyoto 603-8555, Japan}

\author{Chikako Yasui}
\affiliation{National Astronomical Observatory of Japan, 2-21-1 Osawa, Mitaka, Tokyo 181-8588, Japan}
\affiliation{Laboratory of Infrared High-resolution spectroscopy (LiH), Koyama Astronomical Observatory, \\ Kyoto Sangyo University, Motoyama, Kamigamo, Kita-ku, Kyoto 603-8555, Japan}

\author[0000-0003-2161-0361]{Misaki Mizumoto}
\affiliation{Department of Astronomy, Kyoto University, Kitashirakawa-Oiwake-cho, Sakyo-ku, Kyoto 606-8502, Japan}
\affiliation{Hakubi Center, Kyoto University, Yoshida-honmachi, Sakyo-ku, Kyoto 606-8501, Japan}

\author{Shogo Otsubo}
\affiliation{Laboratory of Infrared High-resolution spectroscopy (LiH), Koyama Astronomical Observatory, \\ Kyoto Sangyo University, Motoyama, Kamigamo, Kita-ku, Kyoto 603-8555, Japan}

\author{Keiichi Takenaka}
\affiliation{Department of Physics, Faculty of Science, Kyoto Sangyo University, Motoyama, Kamigamo, Kita-ku, Kyoto 603-8555, Japan}

\author{Tomohiro Yoshikawa}
\affiliation{Edechs, 17203 Iwakura-Minami-Osagi-cho, Sakyo-ku, Kyoto 606-0003, Japan}

\author[0000-0002-9397-3658]{Takuji Tsujimoto}
\affiliation{National Astronomical Observatory of Japan, 2-21-1 Osawa, Mitaka, Tokyo 181-8588, Japan}

%% \nocollaboration{2}



\begin{abstract}
%% To measure detailed chemical abundances of stars, high-resolution spectra mainly in the optical have been used, while the development of near-infrared (near-infrared) spectrographs has opened new wavelength windows.
Red giants show a large number of absorption lines  in both optical and near-infrared wavelengths. Still, the characteristics of the lines in different wave passbands are not necessarily the same.
We searched for lines of \ion{Mg}{1}, \ion{Si}{1}, \ion{Ca}{1}, \ion{Ti}{1}, \ion{Cr}{1}, and \ion{Ni}{1} in the $z^{\prime}$, $Y$, and $J$ bands (0.91--1.33~{$\mu$}m), that are useful for precise abundance analyses, 
from two different compilations of lines, namely, the third release of Vienna Atomic Line Database (VALD3) and the catalog published by Mel{\'e}ndez \& Barbuy in 1999 (MB99).
We selected sufficiently strong lines that are not severely blended and ended up with 191 lines (165 and 141 lines from VALD3 and MB99, respectively), in total, for the six elements.
Combining our line lists with high-resolution ($\lambda/\Delta\lambda = 28,000$) and high signal-to-noise ($>500$) spectra taken with the WINERED spectrograph,
we measured the abundances of the six elements in addition to \ion{Fe}{1} of two prototype red giants, i.e., Arcturus and $\mu$~Leo.
The resultant abundances show reasonable agreements with literature values within {$\sim$}0.2\,dex, 
indicating that the available oscillator strengths are acceptable,
although the abundances based on the two line lists
show systematic differences by 0.1--0.2\,dex.
Furthermore, to improve the precision,
solid estimation of the microturbulence
(or the microturbulences if they are different for different elements) 
is necessary
as far as the classical hydrostatic atmosphere models are used for the analysis.
\end{abstract}



\section{Introduction}
\label{sec:intro}

A list of stellar absorption lines, containing their information such as wavelengths
and oscillator strengths, is essential in the analysis of chemical abundances.
Compared to established lists of lines in the optical, however,
the identification and characterization of absorption lines in the near-infrared range remains incomplete
\citep[see, e.g.,][]{Andreasen-2016,Matsunaga-2020}. 
The correct identification of lines is mandatory, and
estimating the abundances cannot be done accurately without
accurate calibration of the oscillator strengths,
$\loggf$ \footnote{Here and elsewhere in this paper, we consider only the logarithm to base 10.}.

We focus on stellar absorption lines in the $\zYJ$ bands,
0.91--1.33\,{$\mu$}m, in this paper.
In \citet[][hereafter referred to as \citetalias{Kondo-2019}]{Kondo-2019},
we identified 107 \ion{Fe}{1} lines
that are useful for measuring the iron abundances
in the spectra of two well-studied red giants (Arcturus and $\mu$~Leo).
While the iron abundance is one of the most representative parameters that indicates how
stars are chemically enriched, the abundances of other elements
provide us with crucial information on the evolution
of the Milky Way and nearby galaxies \citep{Freeman-2002,Feltzing-2013}. 
Following \citetalias{Kondo-2019},
the purpose of the current study is to extend the identification of
lines in the $\zYJ$ bands to other elements, namely, 
\ion{Mg}{1}, \ion{Si}{1}, \ion{Ca}{1}, \ion{Ti}{1}, \ion{Cr}{1}, and \ion{Ni}{1}.
These elements show {$\sim$}10 or more lines, as we see below,
which would enable precise chemical measurements \citep{Adibekyan-2015}.

In addition to the quality of the line list,
the microturbulence, $\xi$, is a critical ingredient for
performing abundance measurements.
This parameter is not required as long as one deals with weak lines within the linear region of the curve of growth. However, in practice, stronger lines are often included in the analysis to secure a sufficient number of lines. The $\xi$ is required to reproduce the saturated region of the curve of growth
with classical 1D atmospheric models 
with the local thermodynamic equilibrium (LTE) assumed \citep{Gray-2005}, 
while 3D hydrodynamical models do not require $\xi$ given as an external parameter \citep{Asplund-2000,Amarsi-2016}.
The 3D models are also expected to
include naturally the systematic effects caused by 3D/spherical structures
in extended red giants \citep{Dobrovolskas-2013}.
Although the use of 3D hydrodynamical models has been gradually explored
\citep[][and references therein]{Jofre-2019},
it is currently more common to use 1D models for the abundance analysis
because it is easier to use and allows direct comparisons with
previous results based on 1D models. 
In a classical approach with 1D models, 
a depth-independent $\xi$ is estimated by demanding
that the abundances estimated with individual lines show no dependency
on line strengths, e.g., equivalent widths (EWs, denoted as $W$)
or reduced EWs ($W/\lambda$).
This method requires many lines of the same element covering
a wide range of strengths, and \ion{Fe}{1} lines are most often used.
In \citetalias{Kondo-2019}, we performed a bootstrap analysis to determine $\xi$ and
its error by using more than 50 \ion{Fe}{1} lines in the $\zYJ$ bands. 
Among the six elements we add in this paper, \ion{Si}{1} and \ion{Ti}{1} show
many lines enough for doing the same analysis to determine $\xi$,
and we compare the results obtained with these two elements with that obtained with
\ion{Fe}{1}.

\section{Spectral data and line selection} 

\subsection{Observations and data} \label{sec:ObsData}

We investigate the same $\zYJ$-band spectra used in \citetalias{Kondo-2019}. %% Paper~I. 
The spectra of well-studied
red giants (Arcturus and $\mu$~Leo) were collected with
the WINERED cross-dispersed echelle spectrograph \citep{Ikeda-2016}
with the WIDE mode, giving the resolution of $\lambda/\Delta \lambda=28,000$.
We carried out the observation on February 23, 2013,
with the 1.3~m Araki Telescope at Koyama Astronomical Observatory,
Kyoto Sangyo University, Japan (see more details in Table~1 of \citetalias{Kondo-2019}).
The spectrum of a telluric standard, HIP~76267 (A1\,IV),
was used for the correction of telluric absorption lines with the method
described in \citet{Sameshima-2018}.
The wavelength ranges of the three bands in which the telluric lines can be
well-corrected cover 0.91--0.93, 0.96--1.115, and 1.16--1.33~{$\mu$}m.
The continuum of the spectra
was normalized to the unity after the telluric correction. 
The signal-to-noise ratios (S/N) at around 12,500\,{\AA}
are {$\sim$}1,000 before the telluric correction
and are 850 and 720 in the final spectra of Arcturus and $\mu$~Leo, respectively.
The stellar redshifts were corrected so that the absorption lines can be directly compared with those in synthetic spectra in the wavelength scale of standard air at rest.

\begin{figure*}
\includegraphics[clip,width=1\linewidth]{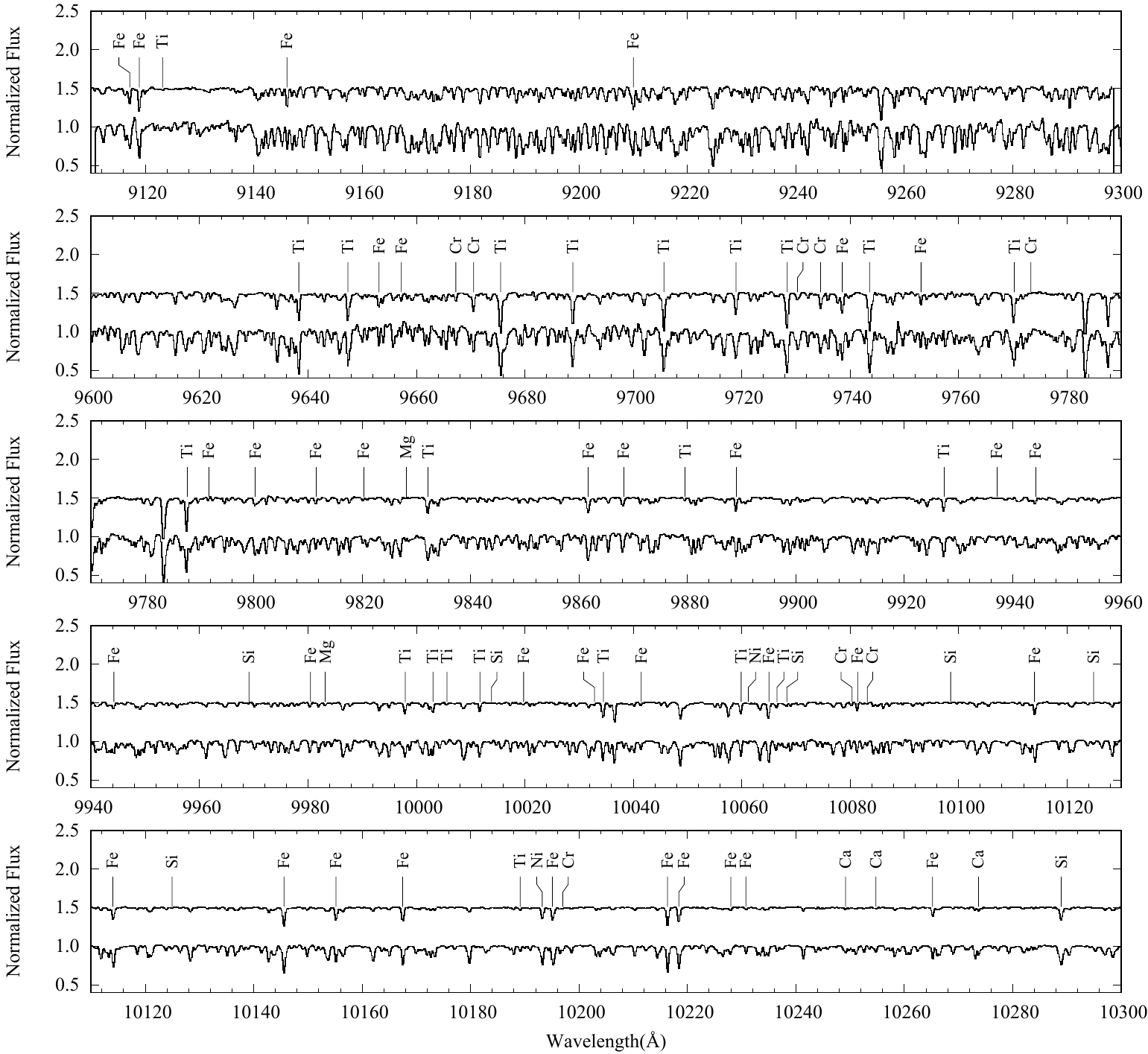}
\caption{
Selected absorption lines seen in Arcturus (upper) and/or $\mu$~Leo (lower).
Their spectra are presented in the scale of wavelength in the standard air.
Tables~\ref{tab1} and \ref{tab2} give details
of the lines selected from VALD3 and MB99, respectively.
\label{fig_spec}}
\end{figure*}

\addtocounter{figure}{-1}
\begin{figure*}
\includegraphics[clip,width=1\linewidth]{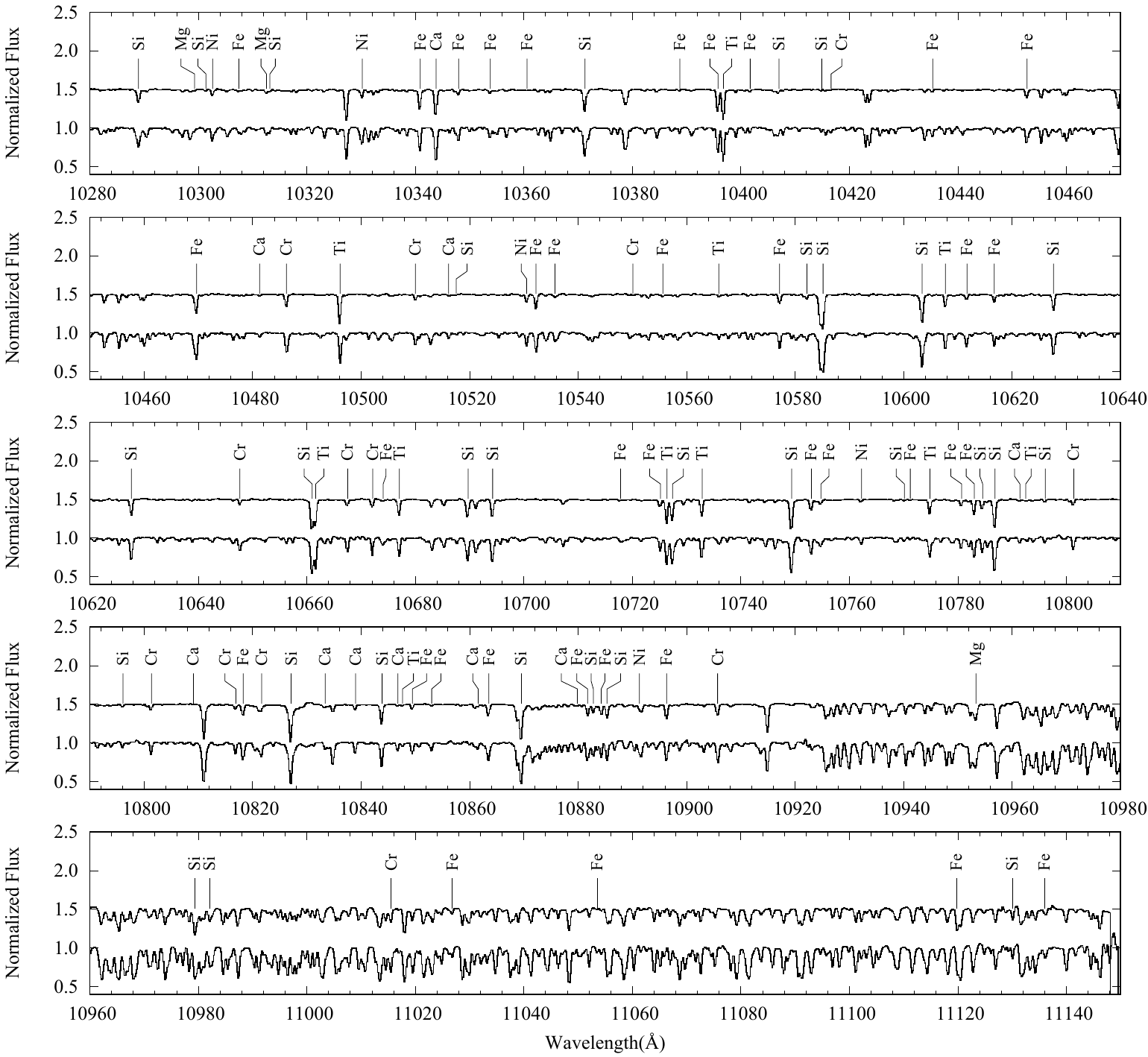}
\caption{(Continued.)}
\end{figure*}

\addtocounter{figure}{-1}
\begin{figure*}
\includegraphics[clip,width=1\linewidth]{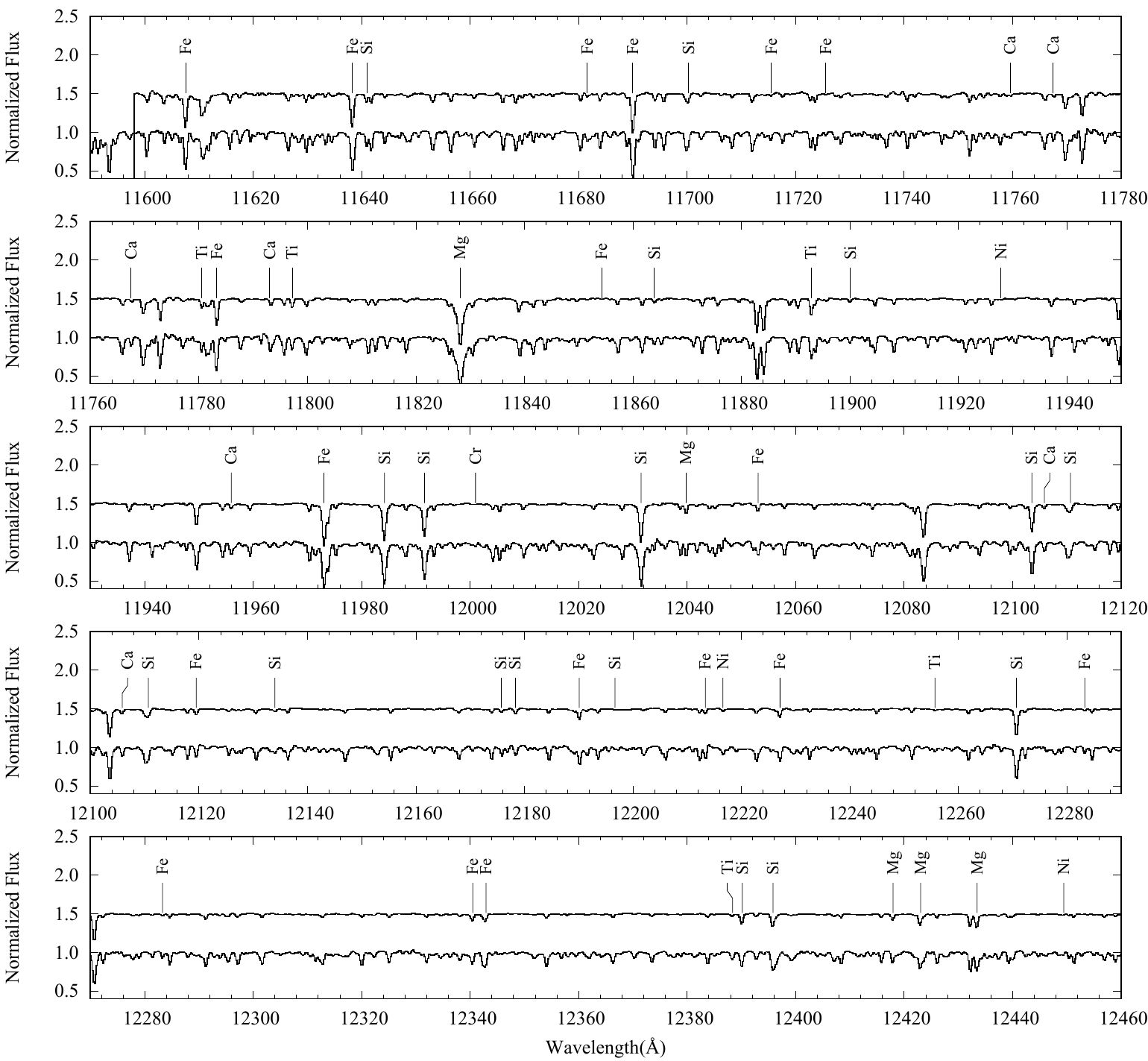}
\caption{(Continued.)}
\end{figure*}

\addtocounter{figure}{-1}
\begin{figure*}
\includegraphics[clip,width=1\linewidth]{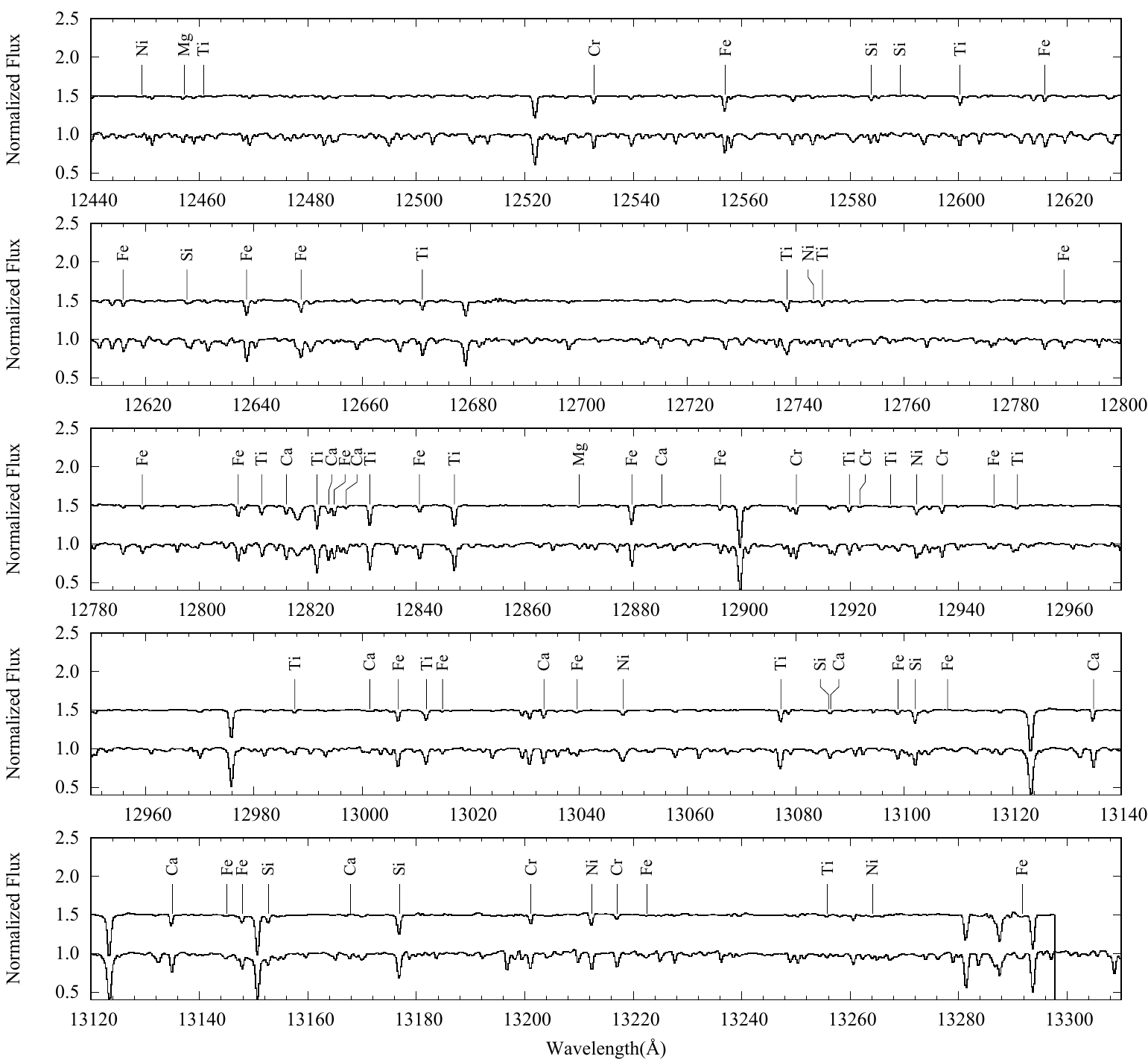}
\caption{(Continued.)}
\end{figure*}

\subsection{Line selection} \label{sec:selection}

We followed the procedure described in \citetalias{Kondo-2019} to select 
the absorption lines that are relatively isolated and useful 
for precise abundance measurements. 
The details that support the following brief description
are given in \citetalias{Kondo-2019}.

We used two line lists,
Vienna Atomic Line Database \citep[VALD3;][]{Ryabchikova-2015}
and the list published by \citet[][hereafter referred to as MB99]{Melendez-1999},
as the starting points of the line selection.
VALD3 has an extensive collection of atomic lines, e.g.,
including more than 10,000 \ion{Fe}{1} lines, and molecular lines
covering the wavelength range of the $\zYJ$ bands.
In our spectrum of Arcturus, Ikeda et~al. (in preparation) identified atomic lines of various species,
including more than 300 lines of \ion{Fe}{1} together with
other elements \citep[see a summary in][Section~3.2]{Taniguchi-2018}.
MB99 used the solar spectrum for the line identification and
the calibration of the $\log gf$ values, and 
compiled {$\sim$}1000 atomic lines in 1.00--1.34\,{$\mu$}m.
The wavelength and the excitation potential ({$\EP$} in eV) of each line are
consistent between the two line lists, but
the $\loggf$ values in the two lists tend to be significantly different.
While we considered
all the lines of the six elements in MB99 as candidates,
the VALD3 lines detected by Ikeda et~al.\  were included, rather than all the VALD3 lines, in the following analysis.

We performed the line selection for the six elements (\ion{Mg}{1}, \ion{Si}{1}, \ion{Ca}{1}, \ion{Ti}{1}, \ion{Cr}{1}, and \ion{Ni}{1})
making use of synthetic spectra
except for the final confirmation with the observed spectra.
In the following analysis, we used the stellar parameters adopted from Heiter et al.\  (2015) as we did in Paper~I; the effective temperature ($\Teff$), the surface gravity ($\log g$), and the global metallicity (${\XH{M}}$) are $4279\pm 40$~K, $1.60\pm 0.18$~dex, and $-0.51\pm 0.06$~dex for Arcturus, and $4520\pm 43$~K, $2.36\pm 0.22$~dex, and $+0.33\pm 0.06$~dex for $\mu$~Leo.
The spectral synthesis was done with SPTOOL developed by Y.~Takeda (private communication), which utilizes the ATLAS9/WIDTH9 codes by R.~L.~Kurucz \citep{Kurucz-1993}.
For each object, we considered two synthetic spectra
for which the atomic lines of VALD3 or MB99 are considered
(i.e.,\ we avoided mixing atomic lines of the two lists in our spectral analysis).
In both cases, we included lines of CN, CO, C$_2$, CH, and OH molecules using the list compiled by VALD3.

As the first step of the line selection, we excluded lines in the following three ranges, as they are severely affected by telluric lines: 9300--9600\,{\AA}, 11150--11600\,{\AA}, and longer than 13300\,{\AA}.
Then, we measured the depths and central wavelengths of the lines in the synthetic spectra for the two objects, Arcturus and $\mu$~Leo.
If the depth of a line
(the distance from the normalized continuum to the line minimum)
was smaller than 0.03 in the synthetic spectra,
the line was rejected.
We also rejected lines with no minimum in the synthetic spectra for
neither of the two objects within 5\,$\kms$ around the expected wavelength.
Besides, when two or more
lines of the same element were detected within 30\,{$\kms$}, we included only the strongest line
if its $X$ value was larger
than those of the other neighboring 
line(s) by more than 0.5\,{dex}; otherwise, we rejected all the lines
in the narrow wavelength range.
The $X$ index is defined as $X \equiv \loggf - \EP \times \theta_{\rm exc}$, where $\theta_{\rm exc} \equiv 5040/(0.86\,\Teff)$.
It is a convenient indicator of line strength \citep{Magain-1984,Gratton-2006}.
These rejections, and also those in the following steps, were made independently for each combination of the line list (VALD3 or MB99) and the object (Arcturus or $\mu$~Leo).

The next step is to evaluate the blending of each target line with neighboring lines
based on a few types of theoretical EWs.
We used two kinds of synthetic spectra generated for each target line, i.e., 
the normal spectrum with all the lines included, $\fsyn$, and 
the one with the target line removed, $\fsynr$.
A normal EW around a target line ($\Lc$) is given by
\begin{equation}
W_i = \int_{\Lc-\Delta_i/2}^{\Lc+\Delta_i/2} \{ 1-\fsyn (\lambda)\} d\lambda,
\label{eq:EW}
\end{equation}
and we considered two different integration ranges, $\Delta_1$ and $\Delta_2$, that correspond to velocities of 30 and 60~$\kms$, respectively.
In addition, we calculated the EW of contaminating lines, $W_i^\dagger$,
which was estimated by Equation~(\ref{eq:EW}) but using $\fsynr$.
Combining these EWs, we consider two indices,
\begin{eqnarray}
\beta_1 &=& W_1^\dagger / W_1 , \\
\beta_2 &=& (W_2^\dagger-W_1^\dagger) / W_1 ,
\end{eqnarray}
as the indicators of blending. The former measures the contamination to the main part
of each target line, and the latter measures the contamination mainly to
the continuum part around the line. We rejected lines with $\beta_1 > 0.3$ or $\beta_2 > 1$
(see Figure~3 in Paper~I for some examples with different $\beta_1$ and $\beta_2$ values).
Finally, we examined whether the lines selected with the synthetic spectra
exist in the observed spectra.
We rejected \ion{Mg}{1}~11820.982 (${\rm EP}=5.933, ~\loggf=-1.520$) selected from VALD3, though not listed in MB99,
because we could not confirm its absorption in the observed spectra. 

\begin{figure}
\begin{center}
\includegraphics[clip,width=1.0\linewidth]{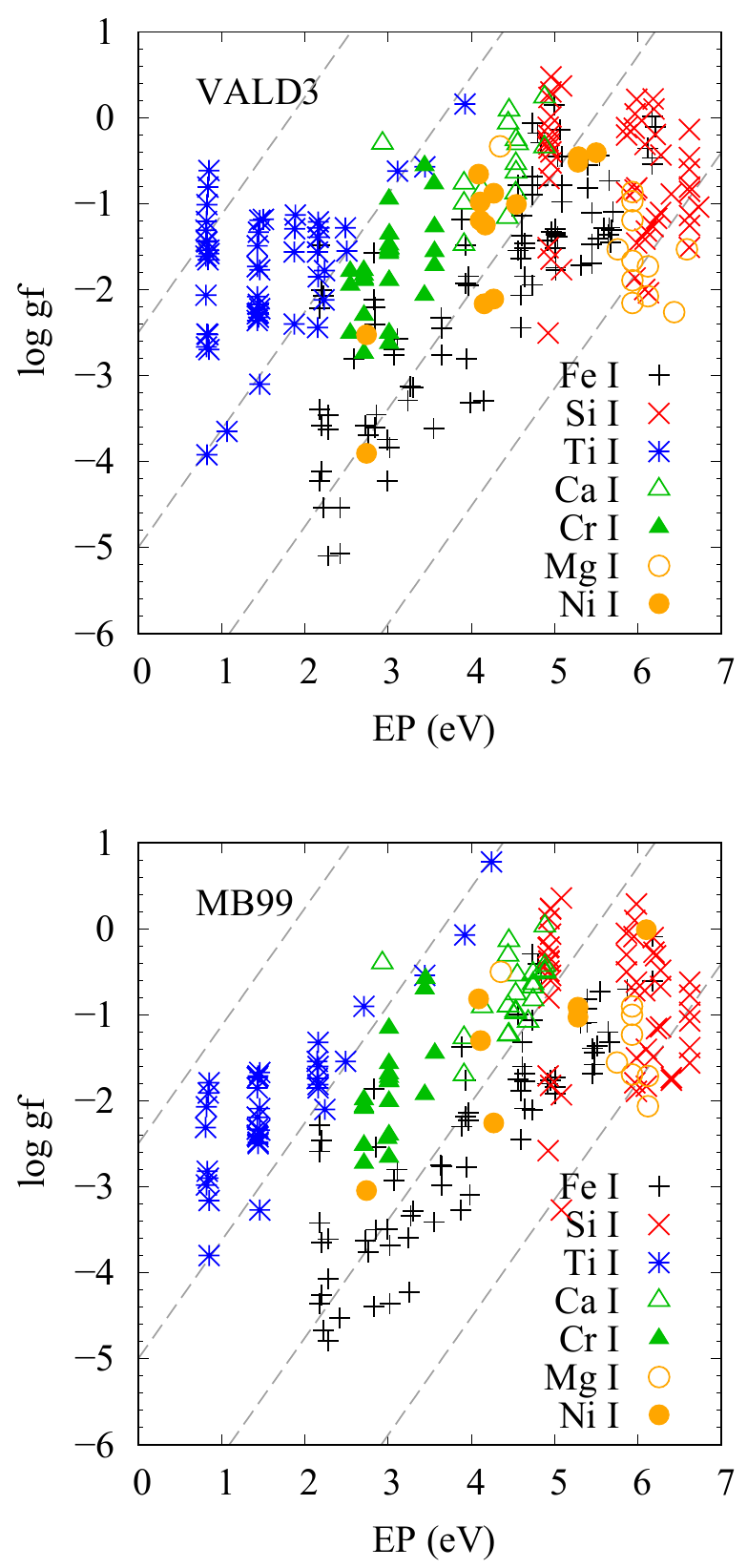}
\end{center}
\caption{
Excitation potential (EP) and the oscillator strength ($\loggf$)
of the selected lines for six elements in addition to Fe.
The lines selected from VALD3 (Table~\ref{tab1}) and those from MB99 (Table~\ref{tab2}) are presented in the upper and lower panels, respectively.
The dashed lines show the contours of the $X$ indicator, $\loggf - \EP \times (5040/(0.86\,\Teff))$, for which $\Teff$ of Arcturus (4286~K) is used.
This indicator nicely predicts which lines are relatively strong 
(stronger towards the upper left corner) among each element,
but the $X$ scale differs from one element to another because
it does not take the elemental abundances into account.
\label{fig_EP-gf}
}
\end{figure}

From VALD3, we selected 12 lines for \ion{Mg}{1}, 50 for \ion{Si}{1}, 15 for \ion{Ca}{1},
50 for \ion{Ti}{1}, 25 for \ion{Cr}{1}, and 13 for \ion{Ni}{1} (Table~\ref{tab1}).
From MB99, the numbers are slightly smaller except for \ion{Ca}{1}:
8 for \ion{Mg}{1}, 49 for \ion{Si}{1}, 22 for \ion{Ca}{1}, 34 for \ion{Ti}{1}, 21 for \ion{Cr}{1}, and 7 for \ion{Ni}{1} (Table~\ref{tab2}).
For \ion{Fe}{1}, 97 and 75 lines, respectively, from VALD3 and MB99
are adopted from \citetalias{Kondo-2019} and included in Tables~\ref{tab1} and \ref{tab2}.
Some lines were selected only for one of the two objects owing to the large difference in metallicity;
some lines were too weak in the metal-poor object, Arcturus, while some other lines
were severely blended with neighboring lines in the metal-rich object, $\mu$~Leo.
Figure~\ref{fig_EP-gf} shows EPs and $\loggf$ values of the selected lines.
The lines of different elements tend to have different EPs;
for example, \ion{Si}{1} have high EPs (${\gtrsim}5$~eV),
\ion{Ti}{1} have low EPs (${\lesssim}3$~eV),
while the EPs of \ion{Fe}{1} lines range from {$\sim$}2 to {$\sim$}6~eV.
The summary of the line selection for individual elements are given in Appendix~\ref{sec:elements}.

\begin{deluxetable}{ccccccc}[!tb]
\tabletypesize{\small}
\tablecaption{Lines Selected from VALD3 and Abundances\label{tab1}}
\tablehead{
  \colhead{$\lambda _{\rm air}$}
& \colhead{Atom}
& \colhead{EP}
& \colhead{$\loggf$}
& \colhead{Arcturus}
& \colhead{$\mu$~Leo}
& \colhead{flag}
\\ 
  \colhead{(\AA)}
& \colhead{}
& \colhead{(eV)}
& \colhead{(dex)}
& \colhead{(dex)}
& \colhead{(dex)}
& \colhead{}
} 
\startdata 
9117.1309 & \ion{Fe}{1} & 2.8581 & $-3.454$ & $-0.37$ & $+0.36$ & 0/2 \\ 
9118.8806 & \ion{Fe}{1} & 2.8316 & $-2.115$ & \multicolumn{1}{c}{(s)} & \multicolumn{1}{c}{(s)} & 0/0 \\ 
9123.2029 & \ion{Ti}{1} & 3.1129 & $-0.619$ & $-0.38$ & $-1.84$: & 1/2 \\ 
9146.1275 & \ion{Fe}{1} & 2.5881 & $-2.804$ & $-0.61$ & $-0.35$ & 0/2 \\ 
9210.0240 & \ion{Fe}{1} & 2.8450 & $-2.404$ & $-0.65$ & \multicolumn{1}{c}{(s)} & 0/1 \\ 
9602.1301 & \ion{Fe}{1} & 5.0117 & $-1.744$ & \multicolumn{1}{c}{(w)} & $+0.11$ & 0/1 \\ 
9638.3043 & \ion{Ti}{1} & 0.8484 & $-0.612$ & \multicolumn{1}{c}{(s)} & \multicolumn{1}{c}{(s)} & 0/0 \\ 
9647.3700 & \ion{Ti}{1} & 0.8181 & $-1.434$ & \multicolumn{1}{c}{(s)} & \multicolumn{1}{c}{(s)} & 0/0 \\ 
9653.1147 & \ion{Fe}{1} & 4.7331 & $-0.684$ & $-0.61$ & $+0.31$ & 0/2 \\ 
9657.2326 & \ion{Fe}{1} & 5.0856 & $-0.780$ & $-0.65$ & $-0.20$ & 0/2 \\ 
\enddata
\tablecomments{
This is the first 10 lines, and the entire list is available as an online material.
The wavelengths, $\lambda _{\rm air}$, are given on the standard air scale.
The abundances, $\XH{X}$, for Arcturus and $\mu$~Leo are scaled
with respect to the solar abundance in \citet{Grevesse-2007},
but the flags, (w), (s), (b), or (*) are given if we did not measure the abundances.
The last column gives the flag of outlier(s), and the colon symbol~(:) accompanies the abundances  that were judged as outliers. See Section 3.1 about the flags.
}
\end{deluxetable}

\begin{deluxetable}{ccccccc}[!tb]
\tabletypesize{\small}
\tablecaption{Lines Selected from MB99 and Abundances\label{tab2}}
\tablehead{
  \colhead{$\lambda _{\rm air}$}
& \colhead{Atom}
& \colhead{EP}
& \colhead{$\loggf$}
& \colhead{Arcturus}
& \colhead{$\mu$~Leo}
& \colhead{flag}
\\ 
  \colhead{(\AA)}
& \colhead{}
& \colhead{(eV)}
& \colhead{(dex)}
& \colhead{(dex)}
& \colhead{(dex)}
& \colhead{}
} 
\startdata 
10003.09 & \ion{Ti}{1} & 2.16 & $-1.32$ & $-0.21$ & $+0.19$ & 0/2 \\ 
10011.74 & \ion{Ti}{1} & 2.15 & $-1.54$ & $-0.03$ & \multicolumn{1}{c}{(b)} & 0/1 \\ 
10013.86 & \ion{Si}{1} & 6.40 & $-1.73$ & \multicolumn{1}{c}{(w)} & $-0.17$ & 0/1 \\ 
10019.79 & \ion{Fe}{1} & 5.48 & $-1.44$ & \multicolumn{1}{c}{(w)} & $+0.30$ & 0/1 \\ 
10032.86 & \ion{Fe}{1} & 5.51 & $-1.36$ & \multicolumn{1}{c}{(w)} & $+0.26$ & 0/1 \\ 
10034.49 & \ion{Ti}{1} & 1.46 & $-2.09$ & $+0.12$ & $+0.65$ & 0/2 \\ 
10041.47 & \ion{Fe}{1} & 5.01 & $-1.84$ & \multicolumn{1}{c}{(w)} & $+0.51$ & 0/1 \\ 
10059.90 & \ion{Ti}{1} & 1.43 & $-2.40$ & $+0.02$ & $+0.54$ & 0/2 \\ 
10065.05 & \ion{Fe}{1} & 4.84 & $-0.57$ & $-0.25$ & $+0.52$ & 0/2 \\ 
10066.55 & \ion{Ti}{1} & 2.16 & $-1.85$ & $-0.10$ & $+0.47$ & 0/2 \\ 
\enddata
\tablecomments{
This is the first 10 lines, and the entire list is available as an online material.
The wavelengths, $\lambda _{\rm air}$, are given on the standard air scale.
The abundances, $\XH{X}$, for Arcturus and $\mu$~Leo are scaled
with respect to the solar abundance in \citet{Grevesse-2007},
but the flags, (w), (s), or (b) are given if we did not measure the abundances.
The last column gives the flag of outlier(s), and the colon symbol~(:) accompanies the abundances  that were judged as outliers. See Section 3.1 about the flags.
}
\end{deluxetable}

\section{Abundance analysis}

\subsection{Measurements of individual lines}
\label{sec:MPFIT}

We measured the abundance by fitting a small spectral part,
within ${\pm}30\,\kms$, around each target line using the MPFIT
tool \citep{Takeda-1995} implemented in SPTOOL
as we did in \citetalias{Kondo-2019}, but we made some changes.
This tool can search for the best match between synthetic and observational spectra
by iteratively varying the abundance
and some other parameters, including the line broadening width.
However, unlike in \citetalias{Kondo-2019}, we fixed the line broadening width
of each spectrum  
to 13.3\,{$\kms$} for Arcturus and 12.0\,{$\kms$} for $\mu$~Leo,
which we determined by measuring the widths of
hundreds of lines of various elements.
These widths correspond to the full-width at the half-maximum
of each absorption line and include the broadenings intrinsic to
the stellar line profile and the instrumental broadening.
As a matter of fact, the instrumental broadening of the WINERED with the WIDE mode is 12\,{$\kms$}, significantly larger than the macroturbulence of Arcturus \citep[5.59\,{$\kms$};][]{Sheminova-2015} and that of $\mu$~Leo \citep[2.9\,{$\kms$};][]{Smith-2013}.
The microturbulence, $\xi$, was fixed at each run of
the MPFIT fitting, and we combined the measurements with different $\xi$
to estimate it (Section~\ref{sec:xieps-Fe}).
Besides, for the spectral synthesis with the MPFIT,
we used the abundance patterns of iron and the six elements
adopted from \citet{Jofre-2015} and
of CNO adopted from \citet{Smith-2013}
for each of our targets (Table~\ref{tab_add}).
When we change the global metallicity, $\XH{M}$, of an atmospheric model,
we increase or decrease $\XH{X}$ by the same amount and
keep these patterns.
This is different from the analysis of \citetalias{Kondo-2019} in which we used
the solar abundance pattern taken from \citet{Anders-1989} for both Arcturus and $\mu$~Leo.
Using the abundance pattern of each object in 
Table~\ref{tab_add}, instead of the solar pattern,
leads to better reproduction
of contaminating lines in each target's spectrum.
When we measure the abundance of
a particular element X, in contrast, we change its $\XH{X}$
but keep the abundances of all the other elements.

\begin{deluxetable}{rcrrrr}[!tb]
\tabletypesize{\small}
\tablecaption{The reference abundances, $\log \epsilon (X)$, used in the MPFIT analysis and as the zero points of $\XH{X}$\label{tab_add}}
\tablehead{
  \colhead{$Z$}
& \colhead{X}
& \twocolhead{Sun}
& \colhead{Arcturus}
& \colhead{$\mu$~Leo}
\\ 
  \colhead{ }
& \colhead{ }
& \colhead{(AG89)}
& \colhead{(G07)}
& \colhead{}
& \colhead{}
} 
\startdata 
6  & C  & 8.56 & 8.39 & 7.96 (S13) & 8.52 (S13) \\
7  & N  & 8.05 & 7.78 & 7.64 (S13) & 8.71 (S13) \\
8  & O  & 8.93 & 8.66 & 8.64 (S13) & 9.05 (S13) \\
12 & Mg & 7.58 & 7.53 & 7.56 (J15) & 8.18 (J15) \\
14 & Si & 7.55 & 7.51 & 7.24 (J15) & 8.02 (J15) \\
20 & Ca & 6.36 & 6.31 & 6.00 (J15) & 6.58 (J15) \\
22 & Ti & 4.99 & 4.90 & 4.56 (J15) & 5.22 (J15) \\
24 & Cr & 5.67 & 5.64 & 5.00 (J15) & 5.93 (J15) \\
26 & Fe & 7.67 & 7.45 & 6.93 (J15) & 7.70 (J15) \\
28 & Ni & 6.25 & 6.23 & 5.68 (J15) & 6.52 (J15) \\
\enddata
\tablerefs{
AG89 = \citet{Anders-1989}; G07 = \citet{Grevesse-2007}; S13 = \citet{Smith-2013}; J15 = \citet{Jofre-2015}
}
\end{deluxetable}

MPFIT gives the abundances in the form of
$\log \epsX{X}=\log N_{\rm X}/\log N_{\rm H}+12$,
where $N_{\rm X}$ indicates the number density of the element ${\rm X}$.
We transformed this form to $\XH{X}=\log \epsX{X}-\log \epsX{X}_\odot$ for which 
we adopted the solar compositions
reported by \citet{Grevesse-2007}
that are given in Table~\ref{tab3}.
The compositions of \citet{Grevesse-2007} were also used
in \citet{Smith-2013} and \citet{Jofre-2015}, with the results of which
we compared our measurements below.

\begin{figure}
\begin{center}
\includegraphics[clip,width=\linewidth]{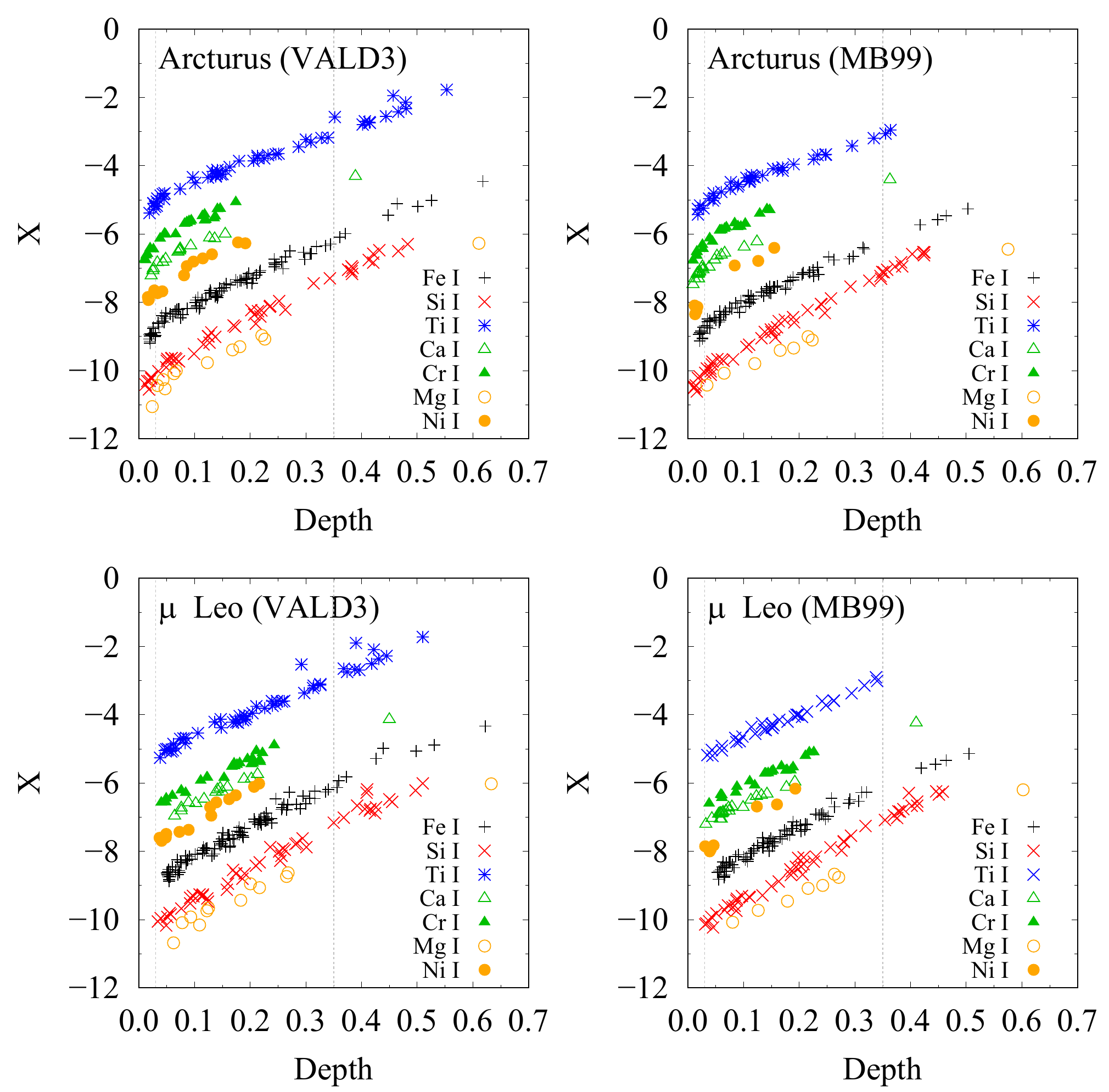}
\end{center}
\caption{
Relationship between the $X$ indices and line depths in the synthetic spectra.
This plot includes all the lines listed in
Table~\ref{tab1} (for VALD3) and Table~\ref{tab2} (for MB99)
regardless of whether they were selected or not for each object
and whether their abundances were measured or not.
The depths here include the absorption by
contaminating lines if any.
In our measurements, the ratio of equivalent width (EW) to depth is
approximately 500\,m{\AA} in the linear region (EW$\lesssim$50\,m{\AA}),
but the depth of 0.35 corresponds to EW$\simeq$190~m{\AA}.
\label{fig:X-depth}
}
\end{figure}

We did not measure the abundances for
the lines that were expected to be too weak,
depth smaller than 0.03, and for those expected to be too much blended
on the basis of $\beta_1$ and $\beta_2$.
The former and latter cases are indicated by the flags (w)
and (b), respectively, in Tables~\ref{tab1} and \ref{tab2}.
Besides, we decided not to measure the abundance for the lines whose depth
is more than 0.35 in the synthetic spectra; the flag of (s) is given to these cases.
In \citetalias{Kondo-2019}, we used the threshold of $X=-6$ for \ion{Fe}{1} lines
for avoiding very strong lines. Since the $X$ indices of
different elements cannot be directly compared, we consider the depth
for this selection, and 0.35 in depth roughly corresponds to 
$X=-6$ in the case of \ion{Fe}{1} (Figure~\ref{fig:X-depth}).
Based on synthesized absorption lines of \ion{Si}{1} and \ion{Ti}{1}, 
we also confirmed that
the damping wing becomes important at the depth of 0.35 or more,
which is consistent with \ion{Fe}{1} lines.
We note that this threshold in depth depends on the spectral resolution
because the resolution of the WINERED is not high enough to resolve
the intrinsic line profile. 
We include the strong lines in our list for completeness and also because
they are expected to be weaker in metal-poor stars. 

While MPFIT failed in Paper~I to measure the abundance of $\mu$~Leo with nine \ion{Fe}{1} lines mainly because of unexpected blends of strong lines, 
we measured the abundances with those \ion{Fe}{1} lines successfully for Arcturus and/or $\mu$~Leo in the new analysis except \ion{Fe}{1}\, 11119.795 which is too strong in both objects.
Some of the lines are contaminated by a neighboring line (or lines) that is
not well reproduced in the synthetic spectra, but fixing the broadening widths
helps to find reasonable fits to the target lines
without being disturbed by the contaminated profile too much. 
On the other hand, we could not measure $\XH{Si}$ of $\mu$~Leo with
\ion{Si}{1}~10407.037 which looks severely disturbed with
the absorption at {$\sim$}10406\,{\AA} that are present 
in the observed spectrum but not predicted by the synthetic spectrum.
The flag of (*) is given to this case in Table~\ref{tab1}.

\subsection{Microturbulence and its effects}
\label{sec:xieps-Fe}

In \citetalias{Kondo-2019}, we applied a bootstrap approach to \ion{Fe}{1} lines
to determine the depth-independent microturbulence ($\xi$)
and the iron abundance
at the same time. In this method, we measure 
the \ion{Fe}{1} abundances with individual lines for a grid of $\xi$ values.
Generally speaking, the estimate of $\XH{X}$ tends to decrease with increasing $\xi$
except for shallow lines whose EWs do not depend on $\xi$. 
The deeper, more saturated lines have the stronger dependency on $\xi$.
Deep lines give higher (or lower) $\XH{X}$ than shallow lines at a lower (higher) $\xi$.
It is naively expected that $\XH{X}$ values with strong/weak lines agree at
the true $\xi$. 
We consider the linear relation, $\XH{X} = a X + b$, where
$X$ is the line strength indicator (Section~\ref{sec:selection}),
to represent
the dependency of $\XH{X}$ on line strength.
The relation, in particular its slope, changes with varying $\xi$.
A common approach for estimating $\xi$ is to find $\xi$ that gives $a=0$
\citep[see, e.g.,][for slightly different approaches]{Blackwell-1977,deJager-1984}.
With the method in \citetalias{Kondo-2019}, 
we generate bootstrapping samples
of the lines, each of which has a sequence of $(\xi, \XH{X})$ values, 
and determine $\xi$ together with $\XH{X}$ for each bootstrapping sample repeatedly.
Then, we evaluate the best estimates of the $(\xi, \XH{X}$) of the star and their errors considering the distribution on the $\xi$--$\XH{X}$ plane 
given by the bootstrapping samples.

\begin{figure*}
\begin{center}
\includegraphics[clip,width=0.8\linewidth]{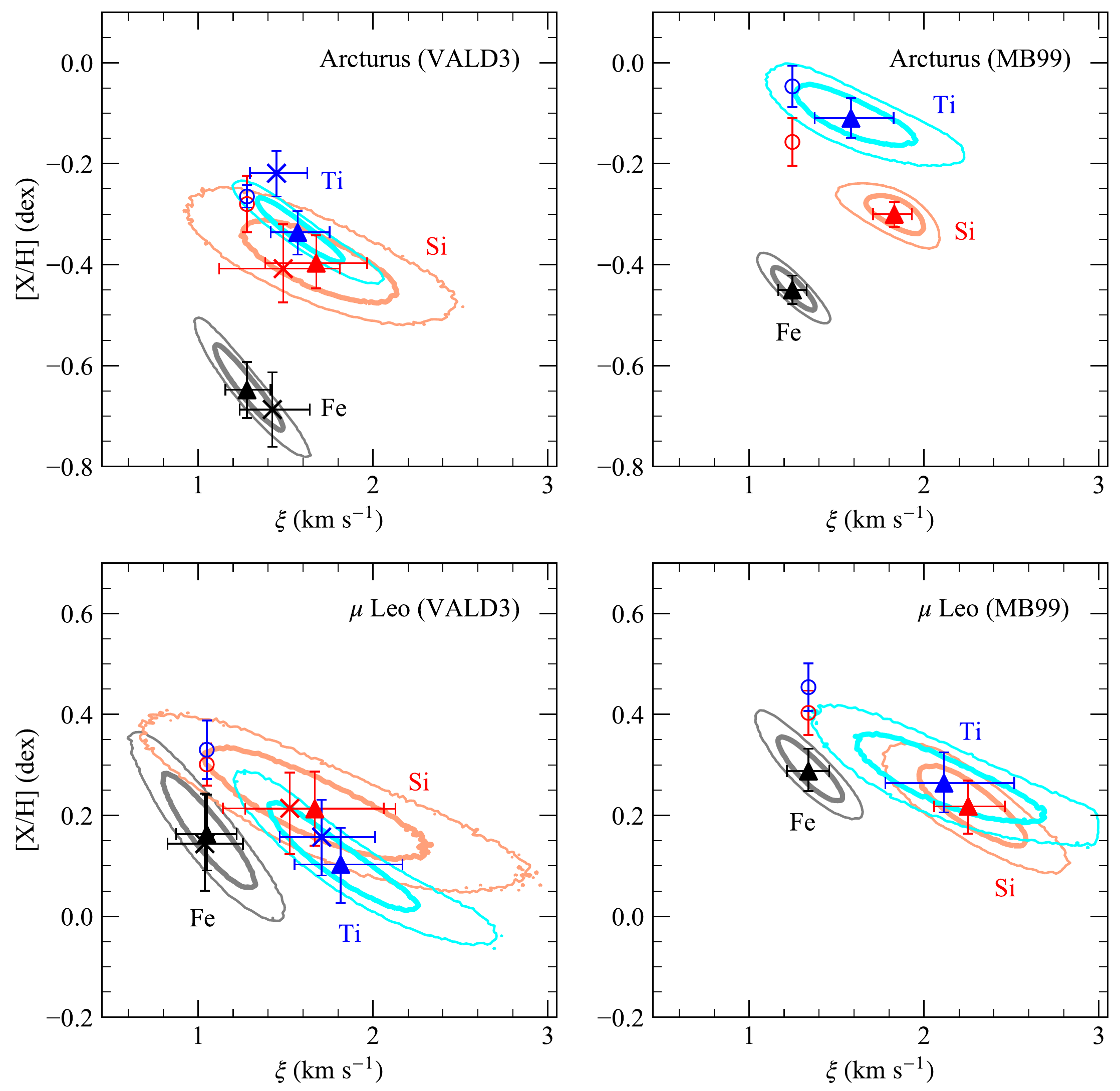}
\end{center}
\caption{
Solutions of the bootstrap method for measuring the abundance, [X/H], and the microturbulence, $\xi$, for the three elements (Fe, Si, and Ti). Each of the four panels shows the results for a combination of the line list (VALD3 or MB99) and the object (Arcturus or $\mu$~Leo).
The best estimates of ($\xi, \XH{X}$) for
individual atoms (listed in Table~\ref{tab3}) are indicated by filled triangles,
and the contours indicate the ranges around the best estimate encircling 68.26\,\% (inner) or 95.44\,\% (outer) of the bootstrap samples.
The best estimates and contours are presented in different colors for different elements: Fe in black, Si in red, and Ti in blue.
For the analysis with VALD, the results with the non-LTE corrections included
(also listed in Table~\ref{tab3}) are indicated by `$\times$' markers.
Open circles for Si and Ti indicate the best estimates with $\xi$ fixed to that obtained for Fe (listed in Table~\ref{tab4}).
}
\label{fig_bootstrap}
\end{figure*}

Since we made some changes in measuring the abundances of individual lines
(Section~\ref{sec:MPFIT}), we performed the bootstrap analysis 
on \ion{Fe}{1} lines again. The resultant $\xi$ and $\XH{Fe}$, listed in Table~\ref{tab3},
are consistent with the previous values in \citetalias{Kondo-2019} within uncertainties,
but the errors get slightly smaller. 
In addition, we made the same analysis for \ion{Si}{1} and \ion{Ti}{1}.
Among the elements we investigate, beside \ion{Fe}{1},  
the numbers of only \ion{Si}{1} and \ion{Ti}{1} lines are large enough
for the bootstrap analysis for determining 
$\xi$ together with the abundance.
As shown in Table~\ref{tab3} and Figure~\ref{fig_bootstrap},
the $\xi$ obtained for Si and Ti lines tend to be higher 
than those for Fe lines. 

\begin{deluxetable}{ccccccc}[!tb]
\tabletypesize{\small}
\tablecaption{Microturbulences and Abundances Obtained with Fe, Si, and Ti Lines\label{tab3}}
\tablehead{
  \colhead{Atom}
& \colhead{List}
& \colhead{NLTE}
& \colhead{$N$}
& \colhead{$\xi$}
& \colhead{$\XH{X}$}
& \colhead{$r$}
\\ 
  \colhead{}
& \colhead{}
& \colhead{}
& \colhead{}
& \colhead{($\kms$)}
& \colhead{(dex)}
& \colhead{}
} 
\startdata 
\multicolumn{7}{c}{{\it Arcturus}} \\ 
\ion{Fe}{1} & MB99 & --- & 53 & $1.25_{-0.08}^{+0.08}$ & $-0.45_{-0.03}^{+0.03}$ & $-0.872$ \\ 
\ion{Fe}{1} & VALD & --- & 66 & $1.28_{-0.12}^{+0.14}$ & $-0.65_{-0.06}^{+0.06}$ & $-0.949$ \\ 
\ion{Fe}{1} & VALD & NLTE & 47 & $1.42_{-0.19}^{+0.21}$ & $-0.69_{-0.07}^{+0.07}$ & $-0.948$ \\ 
\ion{Si}{1} & MB99 & --- & 34 & $1.83_{-0.12}^{+0.10}$ & $-0.30_{-0.03}^{+0.02}$ & $-0.663$ \\ 
\ion{Si}{1} & VALD & --- & 31 & $1.68_{-0.29}^{+0.29}$ & $-0.40_{-0.05}^{+0.06}$ & $-0.739$ \\ 
\ion{Si}{1} & VALD & NLTE & 21 & $1.49_{-0.37}^{+0.32}$ & $-0.41_{-0.07}^{+0.09}$ & $-0.835$ \\ 
\ion{Ti}{1} & MB99 & --- & 25 & $1.58_{-0.21}^{+0.24}$ & $-0.11_{-0.04}^{+0.04}$ & $-0.788$ \\ 
\ion{Ti}{1} & VALD & --- & 34 & $1.57_{-0.15}^{+0.18}$ & $-0.34_{-0.04}^{+0.04}$ & $-0.937$ \\ 
\ion{Ti}{1} & VALD & NLTE & 32 & $1.45_{-0.15}^{+0.18}$ & $-0.22_{-0.05}^{+0.04}$ & $-0.907$ \\ 
\hline
\multicolumn{7}{c}{{\it $\mu$~Leo}} \\ 
\ion{Fe}{1} & MB99 & --- & 67 & $1.34_{-0.12}^{+0.12}$ & $0.29_{-0.04}^{+0.04}$ & $-0.879$ \\ 
\ion{Fe}{1} & VALD & --- & 85 & $1.05_{-0.17}^{+0.17}$ & $0.16_{-0.07}^{+0.08}$ & $-0.909$ \\ 
\ion{Fe}{1} & VALD & NLTE & 55 & $1.04_{-0.21}^{+0.22}$ & $0.14_{-0.09}^{+0.10}$ & $-0.903$ \\ 
\ion{Si}{1} & MB99 & --- & 38 & $2.25_{-0.20}^{+0.21}$ & $0.22_{-0.05}^{+0.05}$ & $-0.820$ \\ 
\ion{Si}{1} & VALD & --- & 33 & $1.67_{-0.40}^{+0.46}$ & $0.21_{-0.07}^{+0.07}$ & $-0.834$ \\ 
\ion{Si}{1} & VALD & NLTE & 19 & $1.52_{-0.38}^{+0.54}$ & $0.21_{-0.09}^{+0.07}$ & $-0.871$ \\ 
\ion{Ti}{1} & MB99 & --- & 29 & $2.11_{-0.34}^{+0.40}$ & $0.26_{-0.06}^{+0.06}$ & $-0.834$ \\ 
\ion{Ti}{1} & VALD & --- & 35 & $1.81_{-0.26}^{+0.35}$ & $0.10_{-0.08}^{+0.07}$ & $-0.921$ \\ 
\ion{Ti}{1} & VALD & NLTE & 30 & $1.71_{-0.24}^{+0.31}$ & $0.16_{-0.08}^{+0.07}$ & $-0.902$ \\ 
\enddata
\tablecomments{
The third column (NLTE) indicates whether the non-LTE correction was applied or not.
The fourth column ($N$) indicates the number of lines used for estimating the microturbulence, $\xi$, and the abundance, $\XH{X}$. 
The $\XH{X}$ are scaled with respect to the solar abundance in \citet{Grevesse-2007}.
The last column ($r$) indicates the correlation coefficient (see the definition in \citetalias{Kondo-2019}) between $\xi$ and $\XH{X}$. 
}
\end{deluxetable}

We performed the same bootstrap analysis
for subsets of the \ion{Fe}{1}, \ion{Si}{1}, and \ion{Ti}{1} lines
to which we could apply the non-LTE correction.
\citet{Bergemann-2012,Bergemann-2013} carried out non-LTE line formation calculations in the atmospheres of red supergiants for Fe, Ti, and Si, and discussed the consequences of non-LTE effects for the $J$-band analysis.
It was found that non-LTE effects are small for $J$-band \ion{Fe}{1} lines, but significant for \ion{Ti}{1} and \ion{Si}{1} lines.
We calculated the non-LTE corrections on the lines of \ion{Fe}{1}, \ion{Si}{1}, and \ion{Ti}{1} using their
website{\footnote{http://nlte.mpia.de/gui-siuAC\_secE.php}}.
Their online tool gives the non-LTE corrections
for the abundances obtained with individual lines
for a given set of stellar parameters. 
We added the corrections to $\XH{X}$ with individual VALD3 lines, 
and performed the same bootstrapping analysis,
but with smaller numbers of lines,
to calculate $\xi$ and $\XH{X}$. 
Not all the lines we selected are included in their tool,
but dozens of the lines could be included in this analysis.
Table~\ref{tab3} and Figure~\ref{fig_bootstrap} present
the resultant $\xi$ and $\XH{X}$.
In the case of Arcturus, $\xi$ gets slightly higher,
but the non-LTE corrections for \ion{Fe}{1} lines are mostly within
0.05\,dex and have little impact on the $\xi$.
In fact, the difference of 0.14\,{$\kms$} with and without
the non-LTE corrections can be explained by the difference
in the lines used for the calculation; using the 47 lines 
to which we could apply the corrections leads
to $\xi \sim 1.42\,{\kms}$ even if we do not apply the corrections.
The non-LTE corrections are as large as 0.15\,dex (or $-0.15$\,dex) for some \ion{Si}{1} and \ion{Ti}{1} lines.
In addition, there are weak correlations between
the line strengths and the non-LTE correlations, 
leading to slightly lower $\xi$, though the impacts on $(\xi, \XH{X})$
are not really significant (Figure~\ref{fig_bootstrap}).
The same analysis can be done with the MB99 lines.
However, the primary purpose of this part is to see the relative impact of
the non-LTE corrections on the microturbulence, and we examined the results with
the non-LTE corrections only for the VALD3 line sets.

\begin{figure}
\begin{center}
\includegraphics[clip, width=\linewidth]{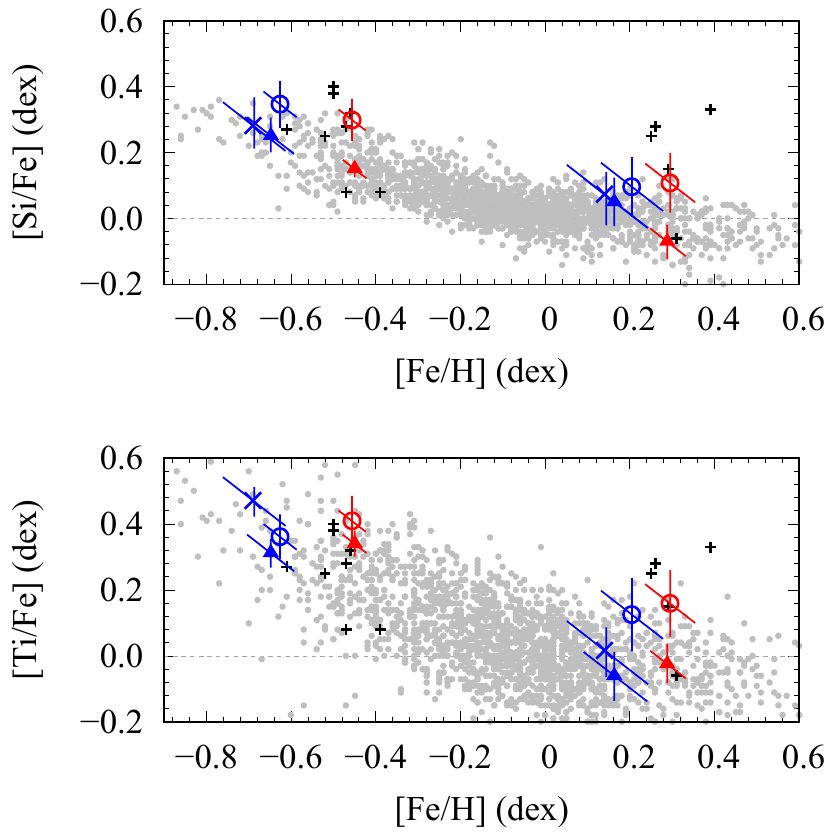}
\end{center}
\caption{
Comparison of the abundances of Arcturus and $\mu$~Leo 
derived by using different methods and line lists
on the $\XFe{X}$--$\FeH$ diagrams.
The results based on the VALD3 lines are illustrated in blue,
and those based on the MB99 lines in red.
The two objects have distinctly different metallicities
(Arcturus being metal poor), and the same markers are used for
both of them within each panel. 
Filled triangles indicate the results with the microturbulences
estimated for individual elements with all the available lines used, while
the VALD3 lines for which the non-LTE effects were taken into account
were used for the results indicated by `$\times$' marker (Table~\ref{tab3}). 
The microturbulences were fixed to those obtained with \ion{Fe}{1} lines
to give the results indicated by open circles (Table~\ref{tab4}).
Small `$+$' markers indicate the literature results for the two objects
(references given in text).
The gray small dots indicate the abundances of red giants
with the APOGEE and Kepler data obtained by \citet{Hawkins-2016}.
Two error bars are given to each point for representing
$\Dtot$ in $\XH{Fe}$ (sloped) and $\Dtot$ in $\XH{X}$ (vertical). 
The total error, $\Dtot$ (Equation~\ref{eq:Dtot}), is explained 
in Section~\ref{sec:abundances}.
%% while the orange and cyan curves indicate the trends
%% of thick-disk and thin-disk stars given by \citet{Bensby-2011}.
\label{fig:FeSiTi}
}
\end{figure}

As the contours in Figure~\ref{fig_bootstrap} suggest,
the abundance we obtain depends on the microturbulence.
If we fix $\xi$ to the one obtained with \ion{Fe}{1} lines,
$\XH{Si}$ and $\XH{Ti}$ get higher than the ones 
obtained with the bootstrap analysis (Section~\ref{sec:abundances}).
If we knew the true abundances, we would be able to determine 
the $\xi$ that leads to accurate abundances. 
However,
we cannot draw such a conclusion because
previously reported values of $\XH{Si}$ and $\XH{Ti}$ have
large scatters. 
Figure~\ref{fig:FeSiTi} compares our estimates with previous results
for Arcturus \citep{Thevenin-1998,Luck-2005,Fulbright-2007,Worley-2009,Chou-2010,Ramirez-2011,Britavskiy-2012,Smith-2013,Jofre-2015}
and $\mu$~Leo \citep{McWilliam-1990,Thevenin-1998,Smith-2000,Luck-2007,Smith-2013,Jofre-2015}.
Here, we only refer to the reports in which
the solar-abundance reference is apparent,
and all the $\XH{X}$ are scaled with respect to
the solar compositions in \citet{Grevesse-2007}.
Regardless of which line lists are used and 
which of the $\xi$ in Table~\ref{tab3} are used,
our estimates are within the scatters of literature values.
%% and also consistent with the trends of disk stars
%% taken from \citet{Bensby-2011}.

\citet{Takeda-1992} investigated the microturbulence of Arcturus and
found that different groups of lines tend to give different $\xi$
showing the correlation with the depth of line forming layers
\citep[also see][for an earlier report on the depth-dependent $\xi$ of Arcturus]{Gray-1981}.
They found that the derived $\xi$ values depend on the properties
of the line sets, e.g., the EPs and the ionization stage of the lines included; $\xi$ tends to be smaller with the lines formed
in the inner atmosphere used in the analysis.
Roughly speaking, the depth of a line forming region is well correlated with the line strength
as expected, while the EP and some other parameters
have additional effects \citep{Gurtovenko-2015}. 
As seen in Figure~\ref{fig:X-depth},
the distributions of depths of the
\ion{Fe}{1}, \ion{Si}{1}, and \ion{Ti}{1} lines used for the analysis
are similarly broad and uniform between the shallowest (0.03)
and the deepest (0.35).
At least, not all the \ion{Si}{1} and \ion{Ti}{1} lines are
formed in layers higher than \ion{Fe}{1} lines. 
There appears to be no simple reason to expect that
both \ion{Si}{1} and \ion{Ti}{1} give $\xi$ higher than \ion{Fe}{1}.
Moreover, while determining $\xi$ accurately requires large numbers
of lines \citep{Mucciarelli-2011}, the numbers of lines 
available for each group tend to be limited.
The problem with the limited numbers of lines is particularly 
severe with infrared spectra; in our case, we can use less than
100 lines even for \ion{Fe}{1}, which results in
the significant statistical errors in $\xi$.
The limited line number prevents us from more detailed discussions such as
using a subset of lines, e.g., those with high or low EPs only, 
for the $\xi$ estimation 
(see the simulation described in Appendix~\ref{sec:bootstrap-simu} 
concerning the number of lines required for the bootstrap analysis).
Moreover, we did not apply the method for determining $\xi$
with the line of the other elements (i.e., \ion{Mg}{1}, \ion{Ca}{1}, and \ion{Ni}{1})
because each of these elements give less than ${\sim}$10 lines.

Considering the above situation, we fix $\xi$ to the one
determined for \ion{Fe}{1} and use it for other elements,
which is a standard
procedure of the classical analysis involved with
the depth-independent $\xi$.
Moreover, the abundances 
measured with the same approach should be used together in discussions about
the features on $\XH{Fe}$--${{\rm [X/Fe]}}$ diagrams, and
it is more common to fix $\xi$ to the one estimated with \ion{Fe}{1} lines. 
The depth dependency of $\xi$ is an important and interesting issue.
Still, the primary purpose of this paper is the line identification,
neither determining the depth-dependent $\xi$ nor 
precise calibration of the oscillator strengths of individual lines.

\subsection{Abundances of the six elements and iron}
\label{sec:abundances}

\begin{figure}
\begin{center}
\includegraphics[clip, width=\linewidth]{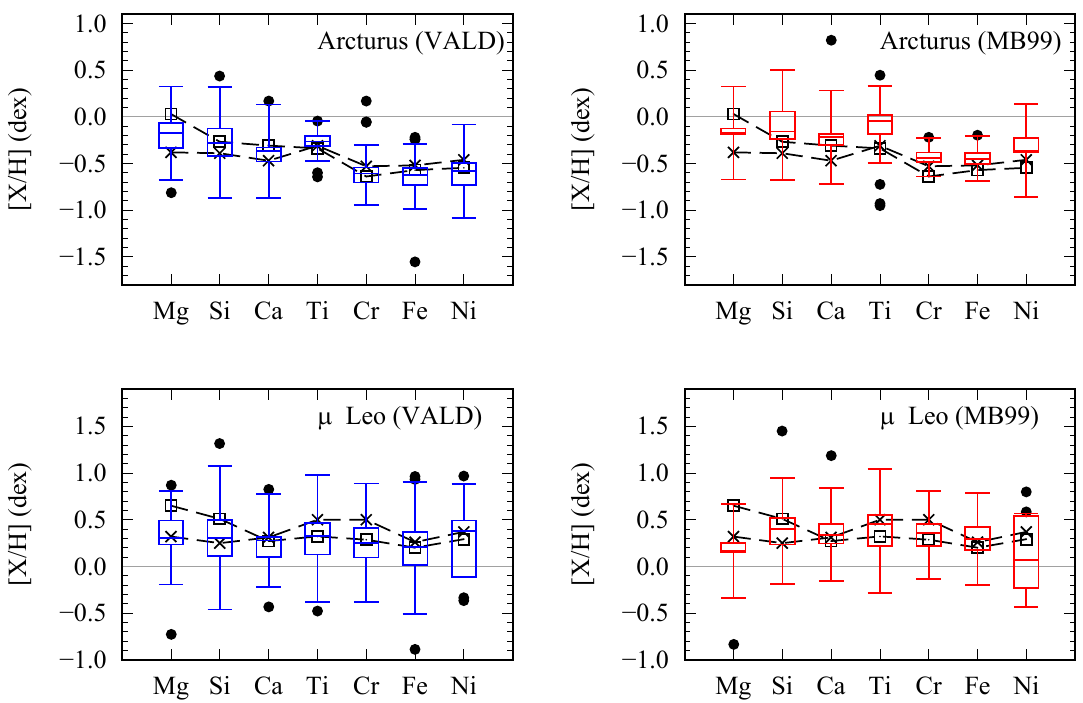}
\end{center}
\caption{
Box plots of abundances derived with individual lines
for each combination of object and line list.
The lower, middle, and upper lines of a box indicate
the quartiles (the 25 percentile, the median, and the 75 percentile).
The upper and lower extremes of the whiskers indicate
the thresholds of the outlier detection,
and black filled circles indicate the outliers (discussed in the text).
Crosses and squares indicate the abundance ratios reported by
\citet{Smith-2013} and 
Jofr\'{e} et~al.\  (\citealt{Jofre-2014} for Fe and \citealt{Jofre-2015} for the other elements), respectively.
\label{fig_box}
}
\end{figure}

In the following analysis, we used the $\xi$ that we
obtained with \ion{Fe}{1} lines without the non-LTE effect corrected
to measure the abundances of 
the six elements (\ion{Mg}{1}, \ion{Si}{1}, \ion{Ca}{1}, \ion{Ti}{1}, \ion{Cr}{1}, and \ion{Ni}{1})
as well as \ion{Fe}{1}.
For each combination of the object (Arcturus and $\mu$~Leo) and the line list (VALD3 or MB99),
we made the MPFIT measurements with the $\xi$ obtained for \ion{Fe}{1}
(Table~\ref{tab3}).
Tables~\ref{tab1} and \ref{tab2} list thus obtained abundances with individual lines selected for each combination.
The flags of (w) or (b) are
given to the lines that were not selected because they are weak or
blended (according to $\beta_1$ and $\beta_2$).
The abundances of the lines with the (s) flag were not 
measured because they are selected but deeper than 0.35 in depth,
while the flag of (*) is given to \ion{Si}{1}~10407.037
in the case of $\mu$~Leo (see Section~\ref{sec:MPFIT}).
Table~\ref{tab4} lists the number, $N$, of the lines
with which we measured $\XH{X}$ 
together with the standard deviation (SD) and the interquartile range (IQR)
of the $N$ values for each combination of line list and object.
Figure~\ref{fig_box} shows box plots of derived abundances.
We took the median of the $N$ values as 
the best estimate of the abundance.
To estimate its error, we took a bootstrap approach;
we generated 10000 bootstrapping samples of the $N$ values and
calculated the SD of the 10000 median values 
from individual bootstrapping samples. 
This SD, $\Dmed$, is considered
as the statistical error of the final $\XH{X}$.
We obtained these estimates for seven elements, including \ion{Fe}{1}
(Table~\ref{tab4}).
It is worthwhile to note that both SDs and IQRs are large for many combinations
of line list and object, ${\rm IQR} \gtrsim 0.15$\,dex for Arcturus and
${\rm IQR} \gtrsim 0.25$\,dex for $\mu$~Leo.
Moreover, the ratio of IQR to SD is expected to be 1.35
for the Gaussian distribution, but the ratios in Table~\ref{tab4} are
not normal for some combinations.
These suggest that the $\log gf$ values in the current lists are not
sufficiently precise, and the errors for some lines are particularly large, 
leading to outliers.
We will come back to this point in Section~\ref{sec:literature}.

\begin{deluxetable}{cccccc}[!tb]
\tabletypesize{\small}
\tablecaption{Abundances Obtained Based on the Two Line Lists\label{tab4}}
\tablehead{
  \colhead{Atom}
& \colhead{List}
& \colhead{$N$}
& \colhead{SD}
& \colhead{IQR}
& \colhead{$\XH{X}$}
\\ 
  \colhead{}
& \colhead{}
& \colhead{}
& \colhead{(dex)}
& \colhead{(dex)}
& \colhead{(dex)}
} 
\startdata 
\multicolumn{6}{c}{{\it Arcturus}}\\
\ion{Mg}{1} & VALD3 & 10 & 0.235 & 0.264  & $-0.176 \pm 0.074 $ \\
\ion{Mg}{1} & MB99 & 6 & 0.100 & 0.056  & $-0.173 \pm 0.042 $ \\
\ion{Si}{1} & VALD3 & 31 & 0.220 & 0.298  & $-0.280 \pm 0.056 $ \\
\ion{Si}{1} & MB99 & 34 & 0.179 & 0.295  & $-0.157 \pm 0.047 $ \\
\ion{Ca}{1} & VALD3 & 11 & 0.177 & 0.149  & $-0.370 \pm 0.040 $ \\
\ion{Ca}{1} & MB99 & 7 & 0.378 & 0.120  & $-0.218 \pm 0.116 $ \\
\ion{Ti}{1} & VALD3 & 34 & 0.120 & 0.107  & $-0.265 \pm 0.022 $ \\
\ion{Ti}{1} & MB99 & 28 & 0.309 & 0.206  & $-0.047 \pm 0.041 $ \\
\ion{Cr}{1} & VALD3 & 18 & 0.262 & 0.161  & $-0.608 \pm 0.032 $ \\
\ion{Cr}{1} & MB99 & 14 & 0.094 & 0.104  & $-0.443 \pm 0.030 $ \\
\ion{Fe}{1} & VALD3 & 66 & 0.196 & 0.174  & $-0.626 \pm 0.018 $ \\
\ion{Fe}{1} & MB99 & 53 & 0.100 & 0.120  & $-0.456 \pm 0.018 $ \\
\ion{Ni}{1} & VALD3 & 9 & 0.151 & 0.232  & $-0.581 \pm 0.070 $ \\
\ion{Ni}{1} & MB99 & 3 & 0.125 & 0.143  & $-0.363 \pm 0.119 $ \\
\hline
\multicolumn{6}{c}{{\it $\mu$~Leo}}\\
\ion{Mg}{1} & VALD3 & 9 & 0.426 & 0.256  & $0.307 \pm 0.122 $ \\
\ion{Mg}{1} & MB99 & 7 & 0.384 & 0.105  & $0.166 \pm 0.115 $ \\
\ion{Si}{1} & VALD3 & 34 & 0.293 & 0.384  & $0.301 \pm 0.042 $ \\
\ion{Si}{1} & MB99 & 36 & 0.313 & 0.284  & $0.403 \pm 0.044 $ \\
\ion{Ca}{1} & VALD3 & 13 & 0.278 & 0.216  & $0.277 \pm 0.054 $ \\
\ion{Ca}{1} & MB99 & 20 & 0.239 & 0.213  & $0.341 \pm 0.034 $ \\
\ion{Ti}{1} & VALD3 & 36 & 0.433 & 0.341  & $0.330 \pm 0.058 $ \\
\ion{Ti}{1} & MB99 & 31 & 0.243 & 0.332  & $0.454 \pm 0.047 $ \\
\ion{Cr}{1} & VALD3 & 25 & 0.234 & 0.316  & $0.248 \pm 0.052 $ \\
\ion{Cr}{1} & MB99 & 21 & 0.203 & 0.236  & $0.361 \pm 0.054 $ \\
\ion{Fe}{1} & VALD3 & 84 & 0.309 & 0.353  & $0.205 \pm 0.033 $ \\
\ion{Fe}{1} & MB99 & 68 & 0.167 & 0.247  & $0.295 \pm 0.027 $ \\
\ion{Ni}{1} & VALD3 & 11 & 0.405 & 0.602  & $0.383 \pm 0.182 $ \\
\ion{Ni}{1} & MB99 & 7 & 0.391 & 0.774  & $0.067 \pm 0.279 $ \\
\enddata
\tablecomments{
For each combination of line list and object, $N$ lines were used for the abundance measurements with the microturbulence, $\xi$, fixed.
The abundances from $N$ individual lines, showing the standard deviation (SD) and the interquartile range (IQR) in the table, were combined to obtain the final abundance, $\XH{X}$, and its error.
}
\end{deluxetable}

\begin{figure}
\begin{center}
\includegraphics[clip, width=0.95\linewidth]{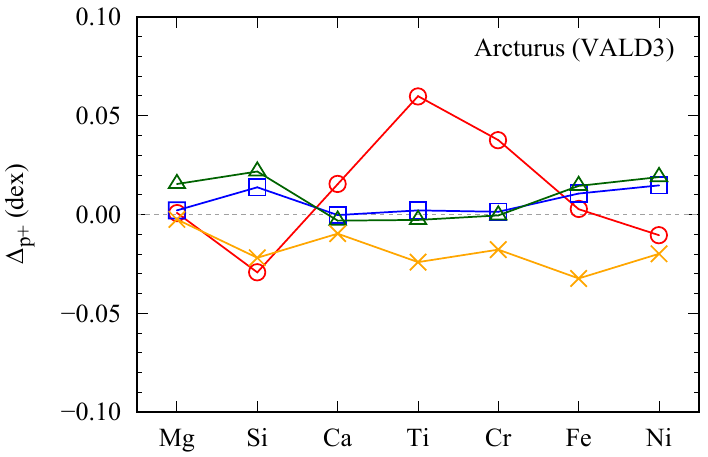}
\end{center}
\caption{
The effects of changing stellar parameters on the abundance of Arcturus measured with the VALD3 lines.
The effects were measured by changing each parameter at a time, and $\Delta_{p+}$ shown in this plot (also see Table~\ref{tab5}) indicates how much $\XH{X}$
changes with the positive offset in the parameter $p$,
$+35$\,K in $\Teff$ (red circles), $+0.06$\,dex in $\log g$ (green triangles),
$+0.08$\,dex in $\XH{M}$ (blue squares), and $+0.14$\,$\kms$ in $\xi$ (orange crosses). 
\label{fig_param}
}
\end{figure}

We also estimated how much the uncertainties in the stellar parameters
($\Teff$, $\log g$, $\XH{M}$, and $\xi$)
affect the estimates of the abundances. 
We adopt the error in each parameter from \citet{Heiter-2015} as
the offset, $\sigma_{p}$, given in Table~\ref{tab5}.
To evaluate the effects of changing these parameters, we added positive and negative offsets to each parameter of the atmosphere models one by one.
For each offset, we ran MPFIT and measured how much the abundance of each line
is altered by the offset.
We then took the median of the relative changes as the impact of each
parameter on the abundance, $\Delta_p$, where $p$ is one of the stellar parameters.
If the sizes of the positive and negative offsets are different from each other,
we consider the root-mean-squares for $\sigma_{p}$ and/or $\Delta_{p}$;
e.g., $\sigma_{\xi}$ tends to be asymmetric, and
$\sigma_{\xi}$ and $\Delta_{\xi}$ in Table~\ref{tab5} are
the root-mean-squares.
Figure~\ref{fig_param} shows the $\Delta_p$ corresponding to
the positive $\sigma_{p}$ offsets of each parameter
for Arcturus with VALD3 used for the measurements.
The size of $\Delta_p$ varies with the element,
and the sign of $\Delta_{p}$ also 
differs from one element to another in the case of $\DT$. 
The trends of $\Delta_p$ seen in Figure~\ref{fig_param}
are similar in the other combinations of line list and object.
The trends for individual elements are described in Appendix~\ref{sec:elements}.

The trend of $\DG$ is noteworthy. This dependency on $\log g$
is very small for \ion{Ca}{1}, \ion{Ti}{1}, and \ion{Cr}{1}, in particular,
while it is non-zero for other elements. For the three elements, the first ionization stage dominates
in the entire range of line-formation layers relevant to this study,
and the line absorption coefficient ($l_\nu$) 
is expected to be proportional to the electron pressure.
The continuum absorption coefficient ($\kappa_\nu$) is also
proportional to the electron pressure around $\Teff$ of our targets  
because it is dominated by the negative hydrogen (H$^{-}$). 
Then, the line depths ($\approx l_\nu/\kappa_\nu$)
are expected to be insensitive to the electron pressure
and to the surface gravity as suggested by the small $\DG$ (\citealt{Gray-2005}; see, also, \citealt{Jian-2020}).
In case of other elements, in contrast, weak lines are formed
in more ionized layers (the ionization fraction {$\sim$}60 to almost 100\,\%),
while the ionization fraction drops significantly,
down to a few percent in some cases,
in the forming layers of strong lines. This results in 
the non-zero dependency on $\log g$. This explains the $\DG$
presented in Figure~\ref{fig_param} well.

Combining the $\Delta_p$ values with the statistical error ($\Dmed$) estimated with the bootstrap analysis,
we calculated the total error, $\Dtot$, by
\begin{eqnarray}
\Dtot^{2} = \DT^{2}+\DG^{2}+\DZ^{2}+\DX^{2}+\Dmed^{2},
\label{eq:Dtot}
\end{eqnarray}
where we ignored the covariant terms
as we did in \citetalias{Kondo-2019}.
The stellar parameters used for Figure~2 of \citetalias{Kondo-2019} show no clear correlation
between any two of them ($\Teff$, $\log g$, $\XH{M}$, and $\xi$).
The results are given in Table~\ref{tab5}.

\begin{figure}
\begin{center}
\includegraphics[clip, width=\linewidth]{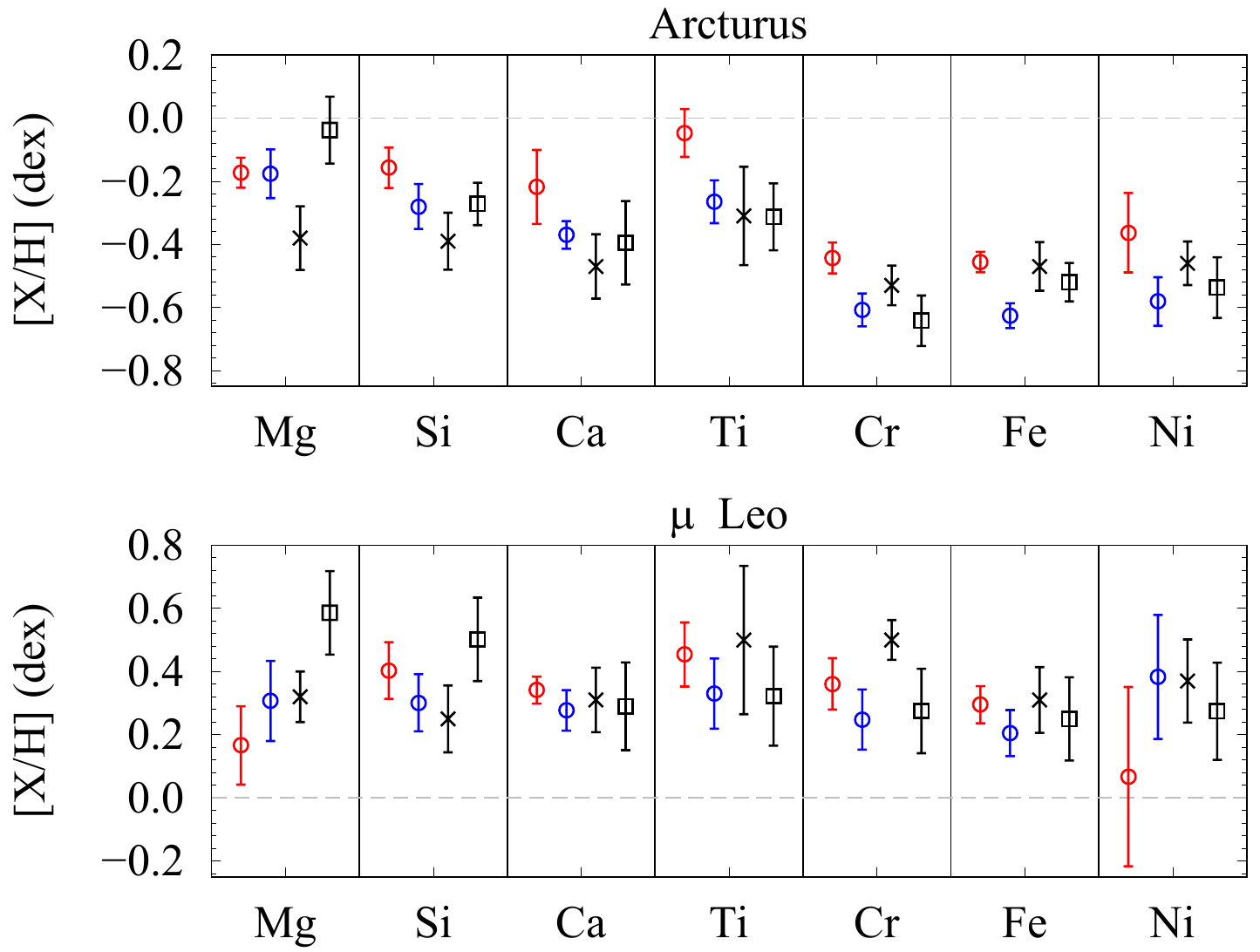}
\end{center}
\caption{
The final abundances of Arcturus and $\mu$~Leo
based on VALD3 (blue) and MB99 (red).
Their statistical errors in Table~\ref{tab4}
and the total errors in Table~\ref{tab5}  
are drawn in the bright colors and pale colors.
Black crosses and squares indicate the values reported by \citet{Smith-2013} and Jofr\'{e} et~al.\  (\citealt{Jofre-2014} for Fe; \citealt{Jofre-2015} for the other elements), respectively.
The definitions of the error bars are given in the text.
\label{fig_median}
}
\end{figure}

\begin{deluxetable*}{lcccccccc}[!tb]
% \tabletypesize{\small}
\tablecaption{Effects of Stellar Parameters and Abundance Errors\label{tab5}}
\tablehead{
\colhead{$\Delta_{p}$}
& \colhead{$\sigma_{p}$}
& \colhead{\ion{Mg}{1}}
& \colhead{\ion{Si}{1}}
& \colhead{\ion{Ca}{1}}
& \colhead{\ion{Ti}{1}}
& \colhead{\ion{Cr}{1}}
& \colhead{\ion{Fe}{1}}
& \colhead{\ion{Ni}{1}}
} 
\startdata 
\multicolumn{9}{c}{{\it Arcturus with VALD3}} \\
$\Delta_{\Teff}$ & $\pm 35$\,K & $\pm 0.001$ & $\mp 0.029$ & $\pm 0.015$ & $\pm 0.060$ & $\pm 0.038$ & $\pm 0.003$ & $\mp 0.011$ \\ 
$\Delta_{\log g}$ & $\pm 0.06$\,dex & $\pm 0.001$ & $\pm 0.014$ & $\mp 0.000$ & $\pm 0.002$ & $\pm 0.001$ & $\pm 0.011$ & $\pm 0.015$ \\ 
$\Delta_{\XH{M}}$ & $\pm 0.08$\,dex & $\pm 0.021$ & $\pm 0.022$ & $\mp 0.003$ & $\mp 0.003$ & $\mp 0.001$ & $\pm 0.014$ & $\pm 0.019$ \\ 
$\Delta_{\xi}$ & $\pm 0.08\,\kms$ & $\mp 0.005$ & $\mp 0.020$ & $\mp 0.009$ & $\mp 0.022$ & $\mp 0.017$ & $\mp 0.030$ & $\mp 0.019$ \\ 
$\Dmed$ & --- & 0.074 & 0.056 & 0.040 & 0.022 & 0.032 & 0.018 & 0.070 \\ 
$\Dtot$ & ---  & 0.077 & 0.071 & 0.044 & 0.068 & 0.052 & 0.039 & 0.077 \\ 
\hline
\multicolumn{9}{c}{{\it Arcturus with MB99}} \\
$\Delta_{\Teff}$ & $\pm 35$\,K & $\pm 0.000$ & $\mp 0.030$ & $\pm 0.015$ & $\pm 0.062$ & $\pm 0.037$ & $\pm 0.008$ & $\mp 0.010$ \\ 
$\Delta_{\log g}$ & $\pm 0.06$\,dex & $\pm 0.003$ & $\pm 0.015$ & $\mp 0.001$ & $\pm 0.003$ & $\pm 0.001$ & $\pm 0.011$ & $\pm 0.017$ \\ 
$\Delta_{\XH{M}}$ & $\pm 0.08$\,dex & $\pm 0.019$ & $\pm 0.023$ & $\mp 0.003$ & $\mp 0.003$ & $\mp 0.001$ & $\pm 0.014$ & $\pm 0.018$ \\ 
$\Delta_{\xi}$ & $\pm 0.13\,\kms$ & $\mp 0.012$ & $\mp 0.016$ & $\mp 0.005$ & $\mp 0.015$ & $\mp 0.011$ & $\mp 0.019$ & $\mp 0.031$ \\ 
$\Dmed$ & --- & 0.042 & 0.047 & 0.116 & 0.041 & 0.030 & 0.018 & 0.119 \\ 
$\Dtot$ & ---  & 0.048 & 0.064 & 0.117 & 0.076 & 0.049 & 0.032 & 0.126 \\ 
\hline
\multicolumn{9}{c}{{\it $\mu$~Leo with VALD3}} \\
$\Delta_{\Teff}$ & $\pm 60$\,K & $\mp 0.008$ & $\mp 0.059$ & $\pm 0.025$ & $\pm 0.082$ & $\pm 0.051$ & $\mp 0.009$ & $\mp 0.026$ \\ 
$\Delta_{\log g}$ & $\pm 0.09$\,dex & $\pm 0.001$ & $\pm 0.021$ & $\mp 0.005$ & $\pm 0.003$ & $\pm 0.004$ & $\pm 0.016$ & $\pm 0.022$ \\ 
$\Delta_{\XH{M}}$ & $\pm 0.15$\,dex & $\pm 0.033$ & $\pm 0.043$ & $\mp 0.006$ & $\pm 0.001$ & $\pm 0.008$ & $\pm 0.032$ & $\pm 0.043$ \\ 
$\Delta_{\xi}$ & $\pm 0.12\,\kms$ & $\mp 0.008$ & $\mp 0.026$ & $\mp 0.023$ & $\mp 0.048$ & $\mp 0.061$ & $\mp 0.053$ & $\mp 0.050$ \\ 
$\Dmed$ & --- & 0.122 & 0.042 & 0.054 & 0.058 & 0.052 & 0.033 & 0.182 \\ 
$\Dtot$ & ---  & 0.127 & 0.090 & 0.064 & 0.111 & 0.095 & 0.073 & 0.197 \\ 
\hline
\multicolumn{9}{c}{{\it $\mu$~Leo with MB99}} \\
$\Delta_{\Teff}$ & $\pm 60$\,K & $\mp 0.007$ & $\mp 0.060$ & $\pm 0.025$ & $\pm 0.085$ & $\pm 0.051$ & $\mp 0.008$ & $\mp 0.022$ \\ 
$\Delta_{\log g}$ & $\pm 0.09$\,dex & $\pm 0.002$ & $\pm 0.025$ & $\mp 0.003$ & $\pm 0.004$ & $\pm 0.004$ & $\pm 0.020$ & $\pm 0.026$ \\ 
$\Delta_{\XH{M}}$ & $\pm 0.15$\,dex & $\pm 0.041$ & $\pm 0.039$ & $\pm 0.001$ & $\pm 0.000$ & $\pm 0.008$ & $\pm 0.031$ & $\pm 0.041$ \\ 
$\Delta_{\xi}$ & $\pm 0.18\,\kms$ & $\mp 0.020$ & $\mp 0.021$ & $\mp 0.006$ & $\mp 0.029$ & $\mp 0.031$ & $\mp 0.036$ & $\mp 0.004$ \\ 
$\Dmed$ & --- & 0.115 & 0.044 & 0.034 & 0.047 & 0.054 & 0.027 & 0.279 \\ 
$\Dtot$ & ---  & 0.124 & 0.090 & 0.043 & 0.101 & 0.081 & 0.059 & 0.284 \\ 
\enddata
\tablecomments{
In the first four lines for each combination of the object and the line list, $\sigma_{p}$ indicates the error, in the parameter $p$, that was used for estimating $\Delta_{p}$, i.e., the effect of $p$ on the abundance $\XH{X}$.
In the last two lines, $\Dmed$ indicates the error in Table~\ref{tab4}, and $\Dtot$ is the total error estimated with Equation~\ref{eq:Dtot}.
}
\end{deluxetable*}

\subsection{Comparison with previous results}
\label{sec:literature}

Figure~\ref{fig_median} compares our two estimates of $\XH{X}$
based on the VALD3 and MB99 for each element
with those reported by 
\citet{Smith-2013} and Jofr{\'e} {et~al.} (\citeyear{Jofre-2014,Jofre-2015}).
The former study used $H$-band spectra, whereas the latter used optical spectra. 
The total errors, $\Dtot$ in Table~\ref{tab5}, are 
used for the error bars for our results.
The errors we use for the results in
\citet{Smith-2013} and Jofr{\'e} {et~al.} also
combine statistical errors and systematic errors.
The definition of the statistical errors given in their works is, however,
different from that of ours. Their statistical errors are
the SDs of abundances from individual lines. 
Their SDs tend to be significantly smaller than the SDs in Table~\ref{tab4}.
This can be ascribed to the difference in the line selection and the quality of $\log gf$ values;
we have not selected the lines on the basis of the accuracy
of $\log gf$. The astrophysical calibration of the $\log gf$ values
remains to be done, desirably, for red giants with each of the abundance-analysis tools.
As for the systematic errors,
all of these works estimated the effects of stellar parameters and, 
for all the points in Figure~\ref{fig_median},
we combined $\Delta_{p}$, in our notation, of
the four parameters ($\Teff$, $\log g$, $\XH{M}$, and $\xi$)
by the square root of sum of squares without
the covariant terms taken into account.

\begin{figure}
\begin{center}
\includegraphics[clip, width=\linewidth]{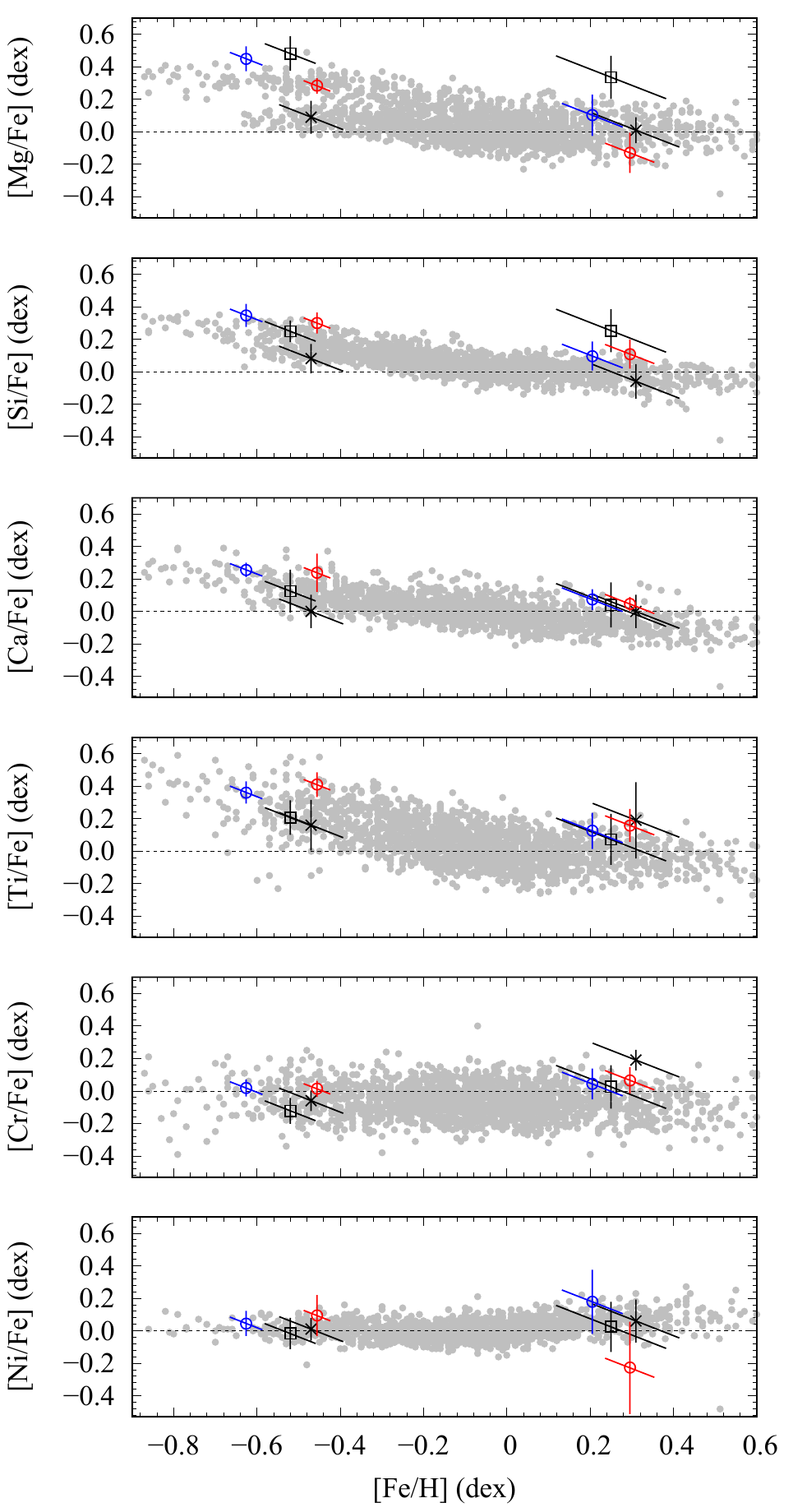}
\end{center}
\caption{
Abundances of Arcturus (metal-poor) and $\mu$~Leo (metal-rich)
derived with different line lists (VALD3 presented in blue and
MB99 in red). The microturbulences were fixed to those obtained
with \ion{Fe}{1} lines for deriving the abundances of all the elements, including Fe. 
Black crosses and squares indicate the values reported by \citet{Smith-2013} and Jofr\'{e} et~al.\  (\citealt{Jofre-2014} for Fe; \citealt{Jofre-2015} for the other elements), respectively.
For each point of $(\XH{Fe}, \XH{X})$,
two error bars are given to represent
the effect of $\Dtot (\XH{Fe})$ (sloped) and that of $\Dtot (\XH{X})$ (vertical). 
The gray small dots indicate the abundances of red giants
with the APOGEE and Kepler data obtained by \citet{Hawkins-2016}.
\label{fig_Fe-X}
}
\end{figure}

Quantitative analysis on the accuracy of $\log gf$, or their re-calibration,
is outside the scope of this paper.
Still, we can identify the lines that seem to have $\log gf$ with
larger errors than other lines, and give caution flags to them. 
\ion{Si}{1}, \ion{Ti}{1}, \ion{Cr}{1}, and \ion{Fe}{1} have
more than 20 lines selected,
and we identified outliers among their lines by considering
the IQR rule.
Let $q_{25}$, $q_{50}$, and $q_{75}$ denote the quartiles
of the $\XH{X}$; $q_{50}$ corresponds to the median,
and the IQR is given by $q_{75}-q_{25}$.
We here use the 1.5$\times$IQR rule for the outlier detection,
i.e., the derived abundance of a line is judged as an outlier if
it is smaller than $q_{25}-1.5\,{\rm IQR}$ or larger than $q_{75}+1.5\,{\rm IQR}$.
In contrast, the numbers of lines are small for \ion{Mg}{1}, \ion{Ca}{1}, and \ion{Ni}{1}, and we use $q_{50}\pm 0.5$\,dex as the thresholds of outliers.
For each line in a given line list, the flag is given as follows.
We combine the number of objects for which the abundance was measured ($N_{\rm obj}$) 
and the number of measurements found to be outliers ($N_{\rm out}$)
and give the flag, $N_{\rm out}/N_{\rm obj}$.
If the line was not used for the abundance measurement,
with the flags such as (w) and (s), for an object,
it is not included in $N_{\rm obj}$
and the outlier detection was not performed.
Thus-determined flags are given in Table~\ref{tab1} and \ref{tab2}. 
The flag of `2/2' suggests that the abundances derived for both objects
are judged as outliers, indicating that the $\loggf$ of such a line has
a large error.
For example, \ion{Ca}{1}~12816.05 was judged as an outlier in all cases,
the $\loggf$ of this line in both VALD3 and MB99 are considered to be rather inaccurate.
We gave non-zero flags (i.e., $N_{\rm out} \neq 0$) to
21 lines in VALD3 and 11 lines in MB99.
We added the colon mark~(:)
to the abundances judged as outliers
in Table~\ref{tab1} and \ref{tab2}.

Concerning the systematics of the abundance scales based on the two line lists, the
$\XH{X}$ based on MB99 are systematically higher
than the counterparts with VALD3 by
0.1--0.2\,dex for both objects except \ion{Mg}{1} and \ion{Ni}{1} for which the numbers of available lines are small. 
Roughly speaking, the best estimates in Figure~\ref{fig_median}
agree with each other allowing for scatters of {$\sim$}0.2\,dex.
Figure~\ref{fig_Fe-X} compares our estimates with
the abundance trends on the $\XFe{X}$--$\XH{Fe}$
diagrams reported by 
\citet{Hawkins-2016}, who measured the abundances of disk stars.
VALD3 and MB99 place Arcturus and $\mu$~Leo
at slightly different positions,
but the results
are still consistent with the trends of disk stars
as expected \citep[e.g.,][]{Smith-2013}.
While the difference of the metallicity scales
is not negligible, $\gtrsim 0.1$\,dex different in $\XH{Fe}$,
we cannot select one of them
based on the comparison in Figure~\ref{fig_Fe-X}.

\section{Concluding remarks}

We identified absorption lines of six elements
(\ion{Mg}{1}, \ion{Si}{1}, \ion{Ca}{1}, \ion{Ti}{1}, \ion{Cr}{1}, and \ion{Ni}{1})
that appear in the $\zYJ$-band spectra of red giants
and are relatively isolated.
From the large compilations of VALD3 and MB99,
{$\sim$}10 (for \ion{Ni}{1} and \ion{Mg}{1}) 
to over 30 (for \ion{Si}{1} and \ion{Ti}{1})
lines were selected for each element.
Some detailed discussions on the selected lines for each element
are given in Appendix~\ref{sec:elements}.
These lines combined with {$>$}50 \ion{Fe}{1} lines reported in
\citetalias{Kondo-2019} are useful for the abundance analysis of
red giants with $\zYJ$-band spectra, which will be
particularly important for obtaining the precise metallicities of
stars obscured by severe interstellar extinction
that hampers optical spectroscopy.

With the abundance analysis making use of 
classical 1D LTE atmospheric models (ATLAS9 in our case),
the microturbulence, $\xi$, is a crucial parameter
required for obtaining accurate results.
We measured $\xi$ that gives $\XH{X}$ independent of line strength
for each of \ion{Si}{1} and \ion{Ti}{1}. They are slightly higher
than those determined for \ion{Fe}{1}, and the differences in
$\xi$ are large enough to give significant impacts on
the abundances, 0.1--0.2\,dex. 
Therefore, unless it turns out that a common $\xi$ can be used for every element, finding a good $\xi$ for each element will be 
an essential step in order to improve the accuracy of the abundance measurements in the future.
Another vital error source is the systematics of the oscillator strengths
given in the two line lists. VALD3 and MB99 tend
to give slightly different abundances, 0.1--0.2\,dex higher if the MB99 is used.
The scatters of the abundances in previous studies
prevent us from judging which of VALD3 and MB99 gives more accurate results. 

\acknowledgments

We are grateful to the staff of Koyama Astronomical Observatory for their support during our observation.
This study is supported by KAKENHI
(Grants-in-Aid, Nos.~26287028 and 18H01248) from the Japan Society for the Promotion of Science (JSPS).
The WINERED was developed by the University of Tokyo and the Laboratory of Infrared High-resolution spectroscopy (LiH), Kyoto Sangyo University 
under the financial supports of JSPS KAKENHI (Nos.~16684001, 20340042, 21840052) and the Supported Programs for the Strategic Research Foundation at Private Universities (Nos~S0801061 and S1411028) from the Ministry of Education, Culture, Sports, Science and Technology (MEXT) of Japan.
This study has made use of the SIMBAD database, operated at CDS, Strasbourg, France and also the VALD database, operated at Uppsala University, the Institute of Astronomy RAS in Moscow, and the University of Vienna.
K.F. is supported by a JSPS Grant-in-Aid for Research Activity Start-up (No.~16H07323).
N.K. is supported by JSPS-DST under the Japan-India Science Cooperative Programs during 2013-2015 and 2016-2018. 

\software
{WINERED} pipeline (S.~Hamano et al. 2021, in preparation),
{IRAF} \citep{Tody-1986,Tody-1993}, 
{PyRAF} (Science Software Branch at STScI, 2012),
{SPTOOL} (Y.~Takeda, private communication).

\appendix

\section{Lines selected for individual elements}
\label{sec:elements}

\begin{enumerate}
\item {\ion{Mg}{1} (Magnesium, $Z=12$)}---
12 and 8 \ion{Mg}{1} lines from VALD3 and MB99, respectively, are included
in our line lists.
\ion{Mg}{1}~11820.982 in VALD3 was listed up with the synthetic spectra
but not included because its absorption was not confirmed in the observed spectra.
All the 8 lines from MB99 are also included in the VALD3 list.
The EPs of 11 lines are high, $>5.9$\,eV, but
\ion{Mg}{1}~11828 has a low EP, {$\sim$}4.35\,eV though it is too strong
to be used for our abundance measurements.
The abundances estimated with individual lines tend to show
a large scatter, but 
the median values are consistent with the literature values
within the errors.
Errors in the stellar parameters have relatively small impacts on 
the measurements of $\XH{Mg}$.

\item {\ion{Si}{1} (Silicon, $Z=14$)}---
56 \ion{Si}{1} lines are included in our lists in total,
50 lines from VALD3 and 49 lines from MB99 (43 lines being selected from both). 
All these lines have relatively high EPs, $\gtrsim 5$\,eV.
Their depths cover a wide range from shallow to deep, and several lines
are too strong to be used for the abundance measurements. 
The abundances with individual lines show large scatters,
but the resultant abundances and the errors are comparable with those in the literature.
$\XH{Si}$ obtained with VALD3 are slightly lower, by ${\sim}$0.1\,dex, than those obtained with MB99.

\item {\ion{Ca}{1} (Calcium, $Z=20$)}---
15 lines are selected from VALD3, while 17 lines are selected from MB99.
There are 10 lines in common, of which \ion{Ca}{1}~10343 in both lists is too strong 
to be used for the abundance measurements. 
More than half of the MB99 lines are too weak in Arcturus but are measurable in $\mu$~Leo.
\ion{Ca}{1}~12816 has highly inaccurate $\loggf$ values, by more than 0.5\,dex, in the two lists,
but other lines lead to reasonable estimates of $\XH{Ca}$.
The abundances estimated with VALD3 are lower than those with MB99
by 0.1--0.15\,dex.

\item {\ion{Ti}{1} (Titanium, $Z=22$)}---
In total, 54 \ion{Ti}{1} lines (50 lines from VALD3 and 34 from MB99)
are included in our lists in total.
Most of them have EPs lower than 2\,eV except a few with $\EP \sim 4$\,eV.
The strengths range from weak to very strong over the limit, 0.35 in depth,
beyond which we did not perform the abundance measurements.
The dependency of derived $\XH{Ti}$ on $\Teff$ and also on $\xi$ is strong. 
The abundances estimated with VALD3 are lower than those with MB99
by 0.1--0.2\,dex.

\item {\ion{Cr}{1} (Chromium, $Z=24$)}---
25 and 21 \ion{Cr}{1} lines are selected from VALD3 and MB99, respectively,
with 19 lines in common and 27 lines in total.
There are not very strong lines, i.e., no \ion{Cr}{1} line is deeper than
0.35 in either Arcturus or $\mu$~Leo, while several lines are too weak in Arcturus. 
The dependency of derived $\XH{Cr}$ on $\Teff$ is relatively strong.
The abundances estimated with VALD3 are lower than those with MB99
by 0.1--0.15\,dex.

\item {\ion{Ni}{1} (Nickel, $Z=28$)}---
15 \ion{Ni}{1} lines (13 and 7 from VALD3 and MB99, respectively)
are included in our line lists.
No line is stronger than the limit of 0.35 in depth, but
a few lines are too weak in Arcturus. The line sets used for
the combinations of line list and object tend to be different.
The abundances from individual lines show large scatters, especially for
$\mu$~Leo, leading to the large errors in the final $\XH{Ni}$, 0.1--0.3\,dex,
combined with the small line numbers. 

\end{enumerate}

\section{Simulations of bootstrap}
\label{sec:bootstrap-simu}

\begin{figure}
\begin{center}
\includegraphics[clip, width=1.0\linewidth]{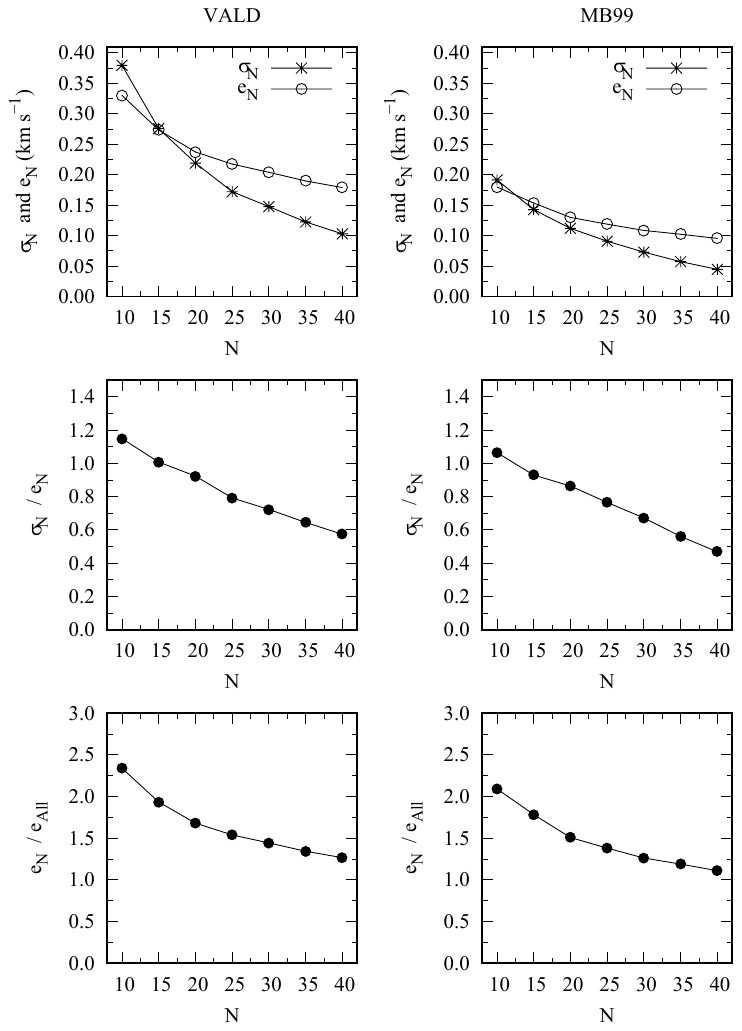}
\end{center}
\caption{
The simulation results on the bootstrap analysis with different numbers, $N$, of lines available.
The parameters are described in the text.
Here presented is the case of Arcturus ($e_{\rm All}=0.14\,\kms$ for VALD3, $0.09\,\kms$ for MB99), and
the left three panels plots the results with VALD3 and the right panels those with MB99.
}
\label{fig_error-V}
\end{figure}

We performed a simulation for investigating how many lines are required 
to derive the microturbulence, $\xi$, and abundance, $\XH{X}$, simultaneously with the bootstrap method.
For this purpose, we used \ion{Fe}{1} lines in Table~\ref{tab1} and \ref{tab2},
but excluded those with flags such as (s) and (w) in either of Arcturus or $\mu$~Leo.
We found 60 VALD3 lines and 50 MB99 lines to use, and
we repeat the following analysis for the four combinations of
object (Arcturus or $\mu$~Leo) and line list (VALD3 or MB99).
Let the set~$A$ indicate this starting line set (the number of lines being $N_{\rm All}=60$ for VALD3 and 50 for MB99).
We first ran the bootstrap analysis for this set~$A$, and compare this result,
$\xi_{\rm All} \pm e_{\rm All}$, with the results with fewer lines as follows.
As simulation sets of fewer lines, we repeated the random selection of $N$ lines
among those in the set~$A$ without duplication 1000 times
for each $N$ ($N$ takes 10, 15, 20, $\cdots$, 40).  
Each set of $N$ lines 
gives an estimate of $\xi(j)$ together with the $1\sigma$ error range,
$[\xi(j)-e_{-}:\xi(j)+e_{+}]$, where $j$ varies from 1 to 1000.
The error in $\xi(j)$ is given by $e(j)=\sqrt{(e_{\xi-}^2+e_{\xi+}^2)/2}$.
We thus obtained 1000 $e(j)$ values and calculated their mean as the measure of
the typical error (denoted as $e_{N}$) obtained with $N$ lines.
In addition, the standard deviation of the 1000 $\xi(j)$, 
denoted as $\sigma_N$, indicates how much the selection of $N$ lines
could affect the resultant microturbulence.
Figure~\ref{fig_error-V} shows these results for Arcturus.
Both $e_{N}$ and $\sigma_N$ increase with decreasing $N$, as expected;
the determination of $\xi$ (and $\XH{X}$) suffers from larger uncertainties
if fewer lines are available. 
When the number of lines gets as small as 15--20,
the statistical uncertainty ($e_{N}$) becomes 1.5--2 times as large as
$e_{\rm All}$. 
In addition, the effect of the line selection ($\sigma_{N}$) becomes
as large as the statistical uncertainty, which means
that the $\xi$ we get is significantly 
affected by the systematic error caused by which lines are
actually available and also by the precision of oscillator strengths
of the particular line set.
%% But there is a tendency for the variance to be large, we need 30 or more lines to determine the parameters.
%% In addition, it should be noted that bootstrap often does not succeed if the number of lines is too small such as $N=10$.
We therefore suggest that the bootstrap method 
of determining $\xi$ and $\XH{X}$ together requires {$\sim$}20 lines at least,
but this limit depends on the precision of $\log gf$.


%\bibliography{sample63}{}
%\bibliographystyle{aasjournal}



% ~\clearpage

\textcolor{white}{\appendix}

\setcounter{table}{0}

\startlongtable
\begin{deluxetable}{ccccccc}%[!tb]
\tabletypesize{\small}
\tablecaption{Lines Selected from VALD3 and Abundances}
\tablehead{
  \colhead{$\lambda _{\rm air}$}
& \colhead{Atom}
& \colhead{EP}
& \colhead{$\loggf$}
& \colhead{Arcturus}
& \colhead{$\mu$~Leo}
& \colhead{flag}
\\ 
  \colhead{(\AA)}
& \colhead{}
& \colhead{(eV)}
& \colhead{(dex)}
& \colhead{(dex)}
& \colhead{(dex)}
& \colhead{}
} 
\startdata 
9117.1309 & \ion{Fe}{1} & 2.8581 & $-$3.454 & $-$0.37 & 0.36 & 0/2 \\
9118.8806 & \ion{Fe}{1} & 2.8316 & $-$2.115 & \multicolumn{1}{c}{(s)} & \multicolumn{1}{c}{(s)} & 0/0 \\
9123.2029 & \ion{Ti}{1} & 3.1129 & $-$0.619 & $-$0.38 & $-$1.84: & 1/2 \\
9146.1275 & \ion{Fe}{1} & 2.5881 & $-$2.804 & $-$0.61 & $-$0.35 & 0/2 \\
9210.0240 & \ion{Fe}{1} & 2.8450 & $-$2.404 & $-$0.65 & \multicolumn{1}{c}{(s)} & 0/1 \\
9602.1301 & \ion{Fe}{1} & 5.0117 & $-$1.744 & \multicolumn{1}{c}{(w)} & 0.11 & 0/1 \\
9638.3043 & \ion{Ti}{1} & 0.8484 & $-$0.612 & \multicolumn{1}{c}{(s)} & \multicolumn{1}{c}{(s)} & 0/0 \\
9647.3700 & \ion{Ti}{1} & 0.8181 & $-$1.434 & \multicolumn{1}{c}{(s)} & \multicolumn{1}{c}{(s)} & 0/0 \\
9653.1147 & \ion{Fe}{1} & 4.7331 & $-$0.684 & $-$0.61 & 0.31 & 0/2 \\
9657.2326 & \ion{Fe}{1} & 5.0856 & $-$0.780 & $-$0.65 & $-$0.20 & 0/2 \\
9667.2850 & \ion{Cr}{1} & 2.5438 & $-$2.510 & $-$0.05 & 0.81: & 1/2 \\
9670.5390 & \ion{Cr}{1} & 2.5443 & $-$1.790 & 0.17 & 0.53: & 1/2 \\
9675.5433 & \ion{Ti}{1} & 0.8360 & $-$0.804 & \multicolumn{1}{c}{(s)} & \multicolumn{1}{c}{(b)} & 0/0 \\
9688.8729 & \ion{Ti}{1} & 0.8129 & $-$1.610 & \multicolumn{1}{c}{(s)} & \multicolumn{1}{c}{(b)} & 0/0 \\
9705.6640 & \ion{Ti}{1} & 0.8259 & $-$1.009 & \multicolumn{1}{c}{(s)} & \multicolumn{1}{c}{(s)} & 0/0 \\
9718.9596 & \ion{Ti}{1} & 1.5025 & $-$1.181 & $-$0.21 & 0.42 & 0/2 \\
9728.4069 & \ion{Ti}{1} & 0.8181 & $-$1.206 & \multicolumn{1}{c}{(s)} & \multicolumn{1}{c}{(s)} & 0/0 \\
9730.3160 & \ion{Cr}{1} & 3.5505 & $-$0.770 & $-$0.58 & 0.65 & 0/2 \\
9734.5630 & \ion{Cr}{1} & 2.5438 & $-$1.950 & $-$0.06 & 0.70: & 1/2 \\
9738.5725 & \ion{Fe}{1} & 4.9913 & 0.150 & $-$0.54 & 0.00 & 0/2 \\
9743.6059 & \ion{Ti}{1} & 0.8129 & $-$1.306 & \multicolumn{1}{c}{(s)} & \multicolumn{1}{c}{(s)} & 0/0 \\
9753.0906 & \ion{Fe}{1} & 4.7955 & $-$0.782 & $-$0.60 & $-$0.30 & 0/2 \\
9770.3011 & \ion{Ti}{1} & 0.8484 & $-$1.581 & \multicolumn{1}{c}{(s)} & \multicolumn{1}{c}{(s)} & 0/0 \\
9773.3530 & \ion{Cr}{1} & 3.5561 & $-$1.559 & \multicolumn{1}{c}{(w)} & $-$0.04 & 0/1 \\
9787.6874 & \ion{Ti}{1} & 0.8259 & $-$1.444 & \multicolumn{1}{c}{(s)} & \multicolumn{1}{c}{(b)} & 0/0 \\
9791.6983 & \ion{Fe}{1} & 2.9904 & $-$4.223 & \multicolumn{1}{c}{(w)} & $-$0.30 & 0/1 \\
9800.3075 & \ion{Fe}{1} & 5.0856 & $-$0.453 & $-$0.96 & $-$0.16 & 0/2 \\
9811.5041 & \ion{Fe}{1} & 5.0117 & $-$1.362 & $-$0.37 & 0.34 & 0/2 \\
9820.2408 & \ion{Fe}{1} & 2.4242 & $-$5.073 & \multicolumn{1}{c}{(w)} & 0.03 & 0/1 \\
9828.1320 & \ion{Mg}{1} & 5.9328 & $-$1.890 & $-$0.81 & 0.71: & 1/2 \\
9832.1397 & \ion{Ti}{1} & 1.8871 & $-$1.130 & $-$0.21 & \multicolumn{1}{c}{(b)} & 0/1 \\
9861.7337 & \ion{Fe}{1} & 5.0638 & $-$0.142 & $-$0.79 & \multicolumn{1}{c}{(b)} & 0/1 \\
9868.1857 & \ion{Fe}{1} & 5.0856 & $-$0.979 & $-$0.40 & 0.67 & 0/2 \\
9879.5830 & \ion{Ti}{1} & 1.8732 & $-$2.400 & $-$0.22 & 0.12 & 0/2 \\
9889.0351 & \ion{Fe}{1} & 5.0331 & $-$0.446 & $-$0.45 & 0.30 & 0/2 \\
9927.3506 & \ion{Ti}{1} & 1.8792 & $-$1.290 & $-$0.30 & 0.41 & 0/2 \\
9937.0898 & \ion{Fe}{1} & 4.5931 & $-$2.442 & \multicolumn{1}{c}{(w)} & 0.12 & 0/1 \\
9944.2065 & \ion{Fe}{1} & 5.0117 & $-$1.338 & $-$0.39 & 0.10 & 0/2 \\
9969.1321 & \ion{Si}{1} & 6.0787 & $-$1.996 & \multicolumn{1}{c}{(w)} & 0.04 & 0/1 \\
9980.4629 & \ion{Fe}{1} & 5.0331 & $-$1.379 & $-$0.52 & 0.53 & 0/2 \\
9983.1880 & \ion{Mg}{1} & 5.9315 & $-$2.153 & $-$0.07 & 0.31 & 0/2 \\
9997.9591 & \ion{Ti}{1} & 1.8732 & $-$1.560 & $-$0.17 & 0.23 & 0/2 \\
10003.088 & \ion{Ti}{1} & 2.1603 & $-$1.210 & $-$0.26 & 0.37 & 0/2 \\
10005.664 & \ion{Ti}{1} & 1.0666 & $-$3.650 & \multicolumn{1}{c}{(w)} & 0.32 & 0/1 \\
10011.744 & \ion{Ti}{1} & 2.1535 & $-$1.390 & $-$0.18 & 0.47 & 0/2 \\
10034.491 & \ion{Ti}{1} & 1.4601 & $-$1.770 & $-$0.22 & 0.47 & 0/2 \\
10041.472 & \ion{Fe}{1} & 5.0117 & $-$1.772 & \multicolumn{1}{c}{(w)} & 0.48 & 0/1 \\
10059.904 & \ion{Ti}{1} & 1.4298 & $-$2.080 & $-$0.30 & 0.33 & 0/2 \\
10061.245 & \ion{Ni}{1} & 5.4943 & $-$0.401 & \multicolumn{1}{c}{(w)} & $-$0.11 & 0/1 \\
10065.045 & \ion{Fe}{1} & 4.8349 & $-$0.289 & $-$0.62 & 0.26 & 0/2 \\
10066.512 & \ion{Ti}{1} & 2.1603 & $-$1.850 & $-$0.09 & 0.50 & 0/2 \\
10068.329 & \ion{Si}{1} & 6.0986 & $-$1.318 & $-$0.34 & 0.04 & 0/2 \\
10080.350 & \ion{Cr}{1} & 3.5558 & $-$1.272 & \multicolumn{1}{c}{(b)} & 0.21 & 0/1 \\
10081.393 & \ion{Fe}{1} & 2.4242 & $-$4.537 & $-$0.41 & 0.31 & 0/2 \\
10083.180 & \ion{Cr}{1} & 3.5561 & $-$1.720 & \multicolumn{1}{c}{(w)} & $-$0.14 & 0/1 \\
10114.014 & \ion{Fe}{1} & 2.7586 & $-$3.692 & $-$0.62 & \multicolumn{1}{c}{(b)} & 0/1 \\
10124.930 & \ion{Si}{1} & 6.1248 & $-$2.035 & \multicolumn{1}{c}{(w)} & 0.50 & 0/1 \\
10145.561 & \ion{Fe}{1} & 4.7955 & $-$0.177 & $-$0.60 & \multicolumn{1}{c}{(b)} & 0/1 \\
10155.162 & \ion{Fe}{1} & 2.1759 & $-$4.226 & $-$0.63 & 0.18 & 0/2 \\
10167.468 & \ion{Fe}{1} & 2.1979 & $-$4.117 & $-$0.55 & 0.35 & 0/2 \\
10189.146 & \ion{Ti}{1} & 1.4601 & $-$3.100 & \multicolumn{1}{c}{(w)} & 0.48 & 0/1 \\
10193.224 & \ion{Ni}{1} & 4.0893 & $-$0.656 & $-$0.58 & 0.43 & 0/2 \\
10195.105 & \ion{Fe}{1} & 2.7275 & $-$3.580 & $-$0.57 & 0.38 & 0/2 \\
10216.313 & \ion{Fe}{1} & 4.7331 & $-$0.063 & $-$0.55 & 0.40 & 0/2 \\
10218.408 & \ion{Fe}{1} & 3.0713 & $-$2.760 & $-$0.57 & 0.58 & 0/2 \\
10227.994 & \ion{Fe}{1} & 6.1189 & $-$0.354 & \multicolumn{1}{c}{(w)} & 0.43 & 0/1 \\
10265.217 & \ion{Fe}{1} & 2.2227 & $-$4.537 & $-$0.61 & 0.07 & 0/2 \\
10273.684 & \ion{Ca}{1} & 4.5347 & $-$0.636 & $-$0.33 & $-$0.44: & 1/2 \\
10288.944 & \ion{Si}{1} & 4.9201 & $-$1.511 & $-$0.54 & 0.26 & 0/2 \\
10299.290 & \ion{Mg}{1} & 6.1182 & $-$2.076 & $-$0.13 & $-$0.73: & 1/2 \\
10302.611 & \ion{Ni}{1} & 4.2661 & $-$0.881 & $-$0.64 & 0.38 & 0/2 \\
10307.454 & \ion{Fe}{1} & 4.5931 & $-$2.067 & $-$0.94 & $-$0.20 & 0/2 \\
10312.531 & \ion{Mg}{1} & 6.1182 & $-$1.730 & 0.00 & 0.49 & 0/2 \\
10313.197 & \ion{Si}{1} & 6.3991 & $-$0.886 & $-$0.52 & 0.32 & 0/2 \\
10330.228 & \ion{Ni}{1} & 4.1054 & $-$0.982 & $-$0.53 & 0.69 & 0/2 \\
10340.885 & \ion{Fe}{1} & 2.1979 & $-$3.577 & $-$0.50 & 0.31 & 0/2 \\
10343.819 & \ion{Ca}{1} & 2.9325 & $-$0.300 & \multicolumn{1}{c}{(s)} & \multicolumn{1}{c}{(s)} & 0/0 \\
10347.965 & \ion{Fe}{1} & 5.3933 & $-$0.551 & $-$0.74 & 0.39 & 0/2 \\
10353.804 & \ion{Fe}{1} & 5.3933 & $-$0.819 & $-$0.69 & 0.16 & 0/2 \\
10360.578 & \ion{Fe}{1} & 5.5188 & $-$1.403 & \multicolumn{1}{c}{(w)} & $-$0.51 & 0/1 \\
10371.263 & \ion{Si}{1} & 4.9296 & $-$0.705 & $-$0.33 & 0.24 & 0/2 \\
10388.744 & \ion{Fe}{1} & 5.4457 & $-$1.468 & \multicolumn{1}{c}{(w)} & 0.38 & 0/1 \\
10395.794 & \ion{Fe}{1} & 2.1759 & $-$3.393 & $-$0.57 & 0.13 & 0/2 \\
10396.802 & \ion{Ti}{1} & 0.8484 & $-$1.539 & \multicolumn{1}{c}{(s)} & \multicolumn{1}{c}{(s)} & 0/0 \\
10407.037 & \ion{Si}{1} & 6.6161 & $-$0.597 & $-$0.5: & \multicolumn{1}{c}{(*)} & 1/1 \\
10414.913 & \ion{Si}{1} & 6.6192 & $-$1.137 & \multicolumn{1}{c}{(w)} & $-$0.26 & 0/1 \\
10416.620 & \ion{Cr}{1} & 3.0128 & $-$2.508 & \multicolumn{1}{c}{(w)} & 0.16 & 0/1 \\
10435.355 & \ion{Fe}{1} & 4.7331 & $-$1.945 & \multicolumn{1}{c}{(w)} & 0.39 & 0/1 \\
10469.652 & \ion{Fe}{1} & 3.8835 & $-$1.184 & $-$0.62 & 0.48 & 0/2 \\
10486.250 & \ion{Cr}{1} & 3.0106 & $-$0.949 & $-$0.62 & 0.42 & 0/2 \\
10496.113 & \ion{Ti}{1} & 0.8360 & $-$1.651 & \multicolumn{1}{c}{(s)} & \multicolumn{1}{c}{(s)} & 0/0 \\
10510.010 & \ion{Cr}{1} & 3.0133 & $-$1.535 & $-$0.71 & 0.08 & 0/2 \\
10517.511 & \ion{Si}{1} & 6.7271 & $-$1.038 & \multicolumn{1}{c}{(w)} & 0.04 & 0/1 \\
10530.514 & \ion{Ni}{1} & 4.1054 & $-$1.189 & $-$0.49 & 0.55 & 0/2 \\
10532.234 & \ion{Fe}{1} & 3.9286 & $-$1.480 & $-$0.64 & 0.19 & 0/2 \\
10535.709 & \ion{Fe}{1} & 6.2057 & $-$0.108 & \multicolumn{1}{c}{(w)} & 0.33 & 0/1 \\
10550.100 & \ion{Cr}{1} & 3.0111 & $-$2.629 & \multicolumn{1}{c}{(w)} & 0.41 & 0/1 \\
10555.649 & \ion{Fe}{1} & 5.4457 & $-$1.108 & \multicolumn{1}{c}{(w)} & $-$0.09 & 0/1 \\
10565.952 & \ion{Ti}{1} & 2.2363 & $-$1.777 & $-$0.6: & 0.07 & 1/2 \\
10577.139 & \ion{Fe}{1} & 3.3014 & $-$3.136 & $-$0.57 & 0.53 & 0/2 \\
10582.160 & \ion{Si}{1} & 6.2227 & $-$1.169 & $-$0.24 & 0.30 & 0/2 \\
10585.141 & \ion{Si}{1} & 4.9538 & 0.012 & \multicolumn{1}{c}{(s)} & \multicolumn{1}{c}{(s)} & 0/0 \\
10603.425 & \ion{Si}{1} & 4.9296 & $-$0.305 & \multicolumn{1}{c}{(s)} & \multicolumn{1}{c}{(s)} & 0/0 \\
10607.718 & \ion{Ti}{1} & 0.8484 & $-$2.697 & $-$0.41 & 0.20 & 0/2 \\
10611.686 & \ion{Fe}{1} & 6.1692 & 0.021 & $-$0.56 & 0.56 & 0/2 \\
10616.721 & \ion{Fe}{1} & 3.2671 & $-$3.127 & $-$0.72 & 0.09 & 0/2 \\
10627.648 & \ion{Si}{1} & 5.8625 & $-$0.866 & 0.44: & 1.32: & 2/2 \\
10647.640 & \ion{Cr}{1} & 3.0106 & $-$1.582 & $-$0.70 & 0.19 & 0/2 \\
10660.973 & \ion{Si}{1} & 4.9201 & $-$0.266 & \multicolumn{1}{c}{(s)} & \multicolumn{1}{c}{(s)} & 0/0 \\
10667.520 & \ion{Cr}{1} & 3.0128 & $-$1.481 & $-$0.79 & 0.11 & 0/2 \\
10672.140 & \ion{Cr}{1} & 3.0133 & $-$1.354 & $-$0.67 & 0.35 & 0/2 \\
10674.070 & \ion{Fe}{1} & 6.1692 & $-$0.466 & \multicolumn{1}{c}{(w)} & 0.42 & 0/1 \\
10677.047 & \ion{Ti}{1} & 0.8360 & $-$2.522 & $-$0.38 & 0.19 & 0/2 \\
10689.716 & \ion{Si}{1} & 5.9537 & $-$0.120 & $-$0.03 & 0.64 & 0/2 \\
10694.251 & \ion{Si}{1} & 5.9639 & 0.048 & $-$0.10 & 0.71 & 0/2 \\
10717.806 & \ion{Fe}{1} & 5.5392 & $-$0.436 & $-$1.55: & $-$0.89: & 2/2 \\
10725.185 & \ion{Fe}{1} & 3.6398 & $-$2.763 & $-$0.73 & 0.19 & 0/2 \\
10726.391 & \ion{Ti}{1} & 0.8129 & $-$2.064 & $-$0.20 & 0.51 & 0/2 \\
10727.406 & \ion{Si}{1} & 5.9840 & 0.217 & $-$0.07 & 0.63 & 0/2 \\
10732.864 & \ion{Ti}{1} & 0.8259 & $-$2.515 & $-$0.27 & 0.47 & 0/2 \\
10749.378 & \ion{Si}{1} & 4.9296 & $-$0.205 & \multicolumn{1}{c}{(s)} & \multicolumn{1}{c}{(s)} & 0/0 \\
10753.004 & \ion{Fe}{1} & 3.9597 & $-$1.845 & $-$0.77 & 0.18 & 0/2 \\
10762.255 & \ion{Ni}{1} & 4.1536 & $-$2.165 & \multicolumn{1}{c}{(w)} & 0.97: & 1/1 \\
10770.134 & \ion{Si}{1} & 6.6192 & $-$1.253 & \multicolumn{1}{c}{(w)} & 0.46 & 0/1 \\
10771.228 & \ion{Fe}{1} & 5.5869 & $-$1.285 & \multicolumn{1}{c}{(w)} & $-$0.24 & 0/1 \\
10774.866 & \ion{Ti}{1} & 0.8181 & $-$2.666 & $-$0.30 & 0.34 & 0/2 \\
10780.694 & \ion{Fe}{1} & 3.2367 & $-$3.289 & $-$0.84 & $-$0.08 & 0/2 \\
10783.050 & \ion{Fe}{1} & 3.1110 & $-$2.567 & $-$0.71 & 0.09 & 0/2 \\
10784.562 & \ion{Si}{1} & 5.9639 & $-$0.839 & $-$0.18 & 0.40 & 0/2 \\
10786.849 & \ion{Si}{1} & 4.9296 & $-$0.303 & \multicolumn{1}{c}{(s)} & \multicolumn{1}{c}{(s)} & 0/0 \\
10796.106 & \ion{Si}{1} & 6.1807 & $-$1.266 & $-$0.41 & 0.30 & 0/2 \\
10801.360 & \ion{Cr}{1} & 3.0111 & $-$1.562 & $-$0.80 & 0.26 & 0/2 \\
10816.910 & \ion{Cr}{1} & 3.0128 & $-$1.894 & $-$0.54 & 0.38 & 0/2 \\
10818.274 & \ion{Fe}{1} & 3.9597 & $-$1.948 & $-$0.67 & 0.48 & 0/2 \\
10821.660 & \ion{Cr}{1} & 3.0133 & $-$1.520 & $-$0.71 & 0.06 & 0/2 \\
10827.088 & \ion{Si}{1} & 4.9538 & 0.302 & \multicolumn{1}{c}{(s)} & \multicolumn{1}{c}{(s)} & 0/0 \\
10838.970 & \ion{Ca}{1} & 4.8775 & 0.238 & $-$0.41 & 0.28 & 0/2 \\
10843.858 & \ion{Si}{1} & 5.8625 & $-$0.112 & 0.11 & 0.80 & 0/2 \\
10847.633 & \ion{Ti}{1} & 0.8259 & $-$3.922 & $-$0.65: & $-$0.48: & 2/2 \\
10861.582 & \ion{Ca}{1} & 4.8767 & $-$0.343 & \multicolumn{1}{c}{(w)} & 0.23 & 0/1 \\
10863.518 & \ion{Fe}{1} & 4.7331 & $-$0.895 & $-$0.67 & 0.24 & 0/2 \\
10869.536 & \ion{Si}{1} & 5.0823 & 0.371 & \multicolumn{1}{c}{(s)} & \multicolumn{1}{c}{(s)} & 0/0 \\
10881.758 & \ion{Fe}{1} & 2.8450 & $-$3.604 & $-$0.44 & 0.59 & 0/2 \\
10882.809 & \ion{Si}{1} & 5.9840 & $-$0.815 & \multicolumn{1}{c}{(b)} & 0.31 & 0/1 \\
10884.262 & \ion{Fe}{1} & 3.9286 & $-$1.925 & $-$0.70 & 0.10 & 0/2 \\
10885.333 & \ion{Si}{1} & 6.1807 & 0.221 & $-$0.37 & 0.40 & 0/2 \\
10891.307 & \ion{Ni}{1} & 4.1672 & $-$1.246 & $-$0.47 & \multicolumn{1}{c}{(b)} & 0/1 \\
10896.299 & \ion{Fe}{1} & 3.0713 & $-$2.694 & $-$0.83 & 0.01 & 0/2 \\
10905.710 & \ion{Cr}{1} & 3.4379 & $-$0.561 & $-$0.52 & 0.31 & 0/2 \\
10953.320 & \ion{Mg}{1} & 5.9315 & $-$0.863 & 0.03 & 0.48 & 0/2 \\
10979.308 & \ion{Si}{1} & 4.9538 & $-$0.524 & $-$0.07 & \multicolumn{1}{c}{(s)} & 0/1 \\
10982.058 & \ion{Si}{1} & 6.1910 & 0.104 & $-$0.42 & 0.17 & 0/2 \\
11026.788 & \ion{Fe}{1} & 3.9433 & $-$2.805 & $-$0.42 & 0.21 & 0/2 \\
11053.512 & \ion{Fe}{1} & 3.9841 & $-$3.311 & \multicolumn{1}{c}{(w)} & 0.23 & 0/1 \\
11119.795 & \ion{Fe}{1} & 2.8450 & $-$2.202 & \multicolumn{1}{c}{(s)} & \multicolumn{1}{c}{(s)} & 0/0 \\
11130.028 & \ion{Si}{1} & 6.2060 & $-$0.194 & $-$0.19 & 0.07 & 0/2 \\
11135.958 & \ion{Fe}{1} & 5.3139 & $-$1.718 & \multicolumn{1}{c}{(w)} & 0.96: & 1/1 \\
11607.572 & \ion{Fe}{1} & 2.1979 & $-$2.009 & \multicolumn{1}{c}{(s)} & \multicolumn{1}{c}{(s)} & 0/0 \\
11638.260 & \ion{Fe}{1} & 2.1759 & $-$2.214 & \multicolumn{1}{c}{(s)} & \multicolumn{1}{c}{(s)} & 0/0 \\
11640.941 & \ion{Si}{1} & 6.2739 & $-$0.432 & $-$0.20 & 0.26 & 0/2 \\
11681.594 & \ion{Fe}{1} & 3.5465 & $-$3.615 & \multicolumn{1}{c}{(w)} & 0.26 & 0/1 \\
11689.972 & \ion{Fe}{1} & 2.2227 & $-$2.068 & \multicolumn{1}{c}{(s)} & \multicolumn{1}{c}{(b)} & 0/0 \\
11715.486 & \ion{Fe}{1} & 5.6424 & $-$1.362 & \multicolumn{1}{c}{(w)} & 0.57 & 0/1 \\
11725.562 & \ion{Fe}{1} & 5.6989 & $-$1.096 & \multicolumn{1}{c}{(w)} & $-$0.23 & 0/1 \\
11759.570 & \ion{Ca}{1} & 4.5313 & $-$0.878 & \multicolumn{1}{c}{(w)} & 0.13 & 0/1 \\
11767.481 & \ion{Ca}{1} & 4.5322 & $-$0.536 & $-$0.48 & 0.28 & 0/2 \\
11780.542 & \ion{Ti}{1} & 1.4432 & $-$2.170 & $-$0.31 & $-$0.02 & 0/2 \\
11783.265 & \ion{Fe}{1} & 2.8316 & $-$1.574 & \multicolumn{1}{c}{(s)} & \multicolumn{1}{c}{(s)} & 0/0 \\
11793.043 & \ion{Ca}{1} & 4.5347 & $-$0.258 & $-$0.22 & 0.55 & 0/2 \\
11797.186 & \ion{Ti}{1} & 1.4298 & $-$2.280 & $-$0.31 & 0.17 & 0/2 \\
11828.171 & \ion{Mg}{1} & 4.3458 & $-$0.333 & \multicolumn{1}{c}{(s)} & \multicolumn{1}{c}{(s)} & 0/0 \\
11854.238 & \ion{Fe}{1} & 5.6826 & $-$1.306 & \multicolumn{1}{c}{(w)} & $-$0.23 & 0/1 \\
11863.920 & \ion{Si}{1} & 5.9840 & $-$1.457 & $-$0.28 & 0.30 & 0/2 \\
11892.877 & \ion{Ti}{1} & 1.4298 & $-$1.730 & $-$0.13 & \multicolumn{1}{c}{(b)} & 0/1 \\
11900.055 & \ion{Si}{1} & 5.9639 & $-$1.864 & $-$0.03 & 0.57 & 0/2 \\
11927.838 & \ion{Ni}{1} & 4.2661 & $-$2.105 & \multicolumn{1}{c}{(w)} & $-$0.33 & 1/1 \\
11955.955 & \ion{Ca}{1} & 4.1308 & $-$0.849 & $-$0.37 & 0.29 & 0/2 \\
11973.046 & \ion{Fe}{1} & 2.1759 & $-$1.483 & \multicolumn{1}{c}{(s)} & \multicolumn{1}{c}{(s)} & 0/0 \\
11984.198 & \ion{Si}{1} & 4.9296 & 0.239 & \multicolumn{1}{c}{(s)} & \multicolumn{1}{c}{(s)} & 0/0 \\
11991.568 & \ion{Si}{1} & 4.9201 & $-$0.109 & \multicolumn{1}{c}{(s)} & \multicolumn{1}{c}{(s)} & 0/0 \\
12000.970 & \ion{Cr}{1} & 3.4348 & $-$2.068 & \multicolumn{1}{c}{(w)} & 0.06 & 0/1 \\
12031.504 & \ion{Si}{1} & 4.9538 & 0.477 & \multicolumn{1}{c}{(s)} & \multicolumn{1}{c}{(s)} & 0/0 \\
12039.822 & \ion{Mg}{1} & 5.7532 & $-$1.530 & $-$0.14 & 0.21 & 0/2 \\
12053.082 & \ion{Fe}{1} & 4.5585 & $-$1.543 & $-$0.63 & 0.05 & 0/2 \\
12103.534 & \ion{Si}{1} & 4.9296 & $-$0.350 & \multicolumn{1}{c}{(s)} & \multicolumn{1}{c}{(s)} & 0/0 \\
12105.841 & \ion{Ca}{1} & 4.5541 & $-$0.305 & $-$0.48 & 0.06 & 0/2 \\
12110.659 & \ion{Si}{1} & 6.6161 & $-$0.136 & $-$0.53 & 0.38 & 0/2 \\
12119.494 & \ion{Fe}{1} & 4.5931 & $-$1.635 & $-$0.69 & 0.18 & 0/2 \\
12175.733 & \ion{Si}{1} & 6.6192 & $-$0.855 & $-$0.28 & 0.41 & 0/2 \\
12178.339 & \ion{Si}{1} & 6.2694 & $-$1.100 & $-$0.15 & 0.56 & 0/2 \\
12190.098 & \ion{Fe}{1} & 3.6352 & $-$2.330 & $-$0.79 & 0.22 & 0/2 \\
12213.336 & \ion{Fe}{1} & 4.6382 & $-$1.845 & $-$0.55 & 0.34 & 0/2 \\
12216.579 & \ion{Ni}{1} & 5.2825 & $-$0.513 & $-$0.46 & 0.41 & 0/2 \\
12227.112 & \ion{Fe}{1} & 4.6070 & $-$1.368 & $-$0.80 & $-$0.01 & 0/2 \\
12255.699 & \ion{Ti}{1} & 3.9215 & 0.161 & \multicolumn{1}{c}{(w)} & $-$0.04 & 0/1 \\
12270.692 & \ion{Si}{1} & 4.9538 & $-$0.396 & \multicolumn{1}{c}{(s)} & \multicolumn{1}{c}{(s)} & 0/0 \\
12283.298 & \ion{Fe}{1} & 6.1692 & $-$0.537 & \multicolumn{1}{c}{(w)} & 0.27 & 0/1 \\
12340.481 & \ion{Fe}{1} & 2.2786 & $-$5.098 & $-$0.24 & 0.55 & 1/2 \\
12342.916 & \ion{Fe}{1} & 4.6382 & $-$1.463 & $-$0.71 & 0.36 & 0/2 \\
12390.154 & \ion{Si}{1} & 5.0823 & $-$1.767 & $-$0.42 & 0.37 & 0/2 \\
12395.832 & \ion{Si}{1} & 4.9538 & $-$1.644 & $-$0.44 & 0.28 & 0/2 \\
12417.937 & \ion{Mg}{1} & 5.9315 & $-$1.664 & $-$0.21 & 0.26 & 0/2 \\
12423.029 & \ion{Mg}{1} & 5.9320 & $-$1.188 & $-$0.25 & \multicolumn{1}{c}{(b)} & 0/1 \\
12433.452 & \ion{Mg}{1} & 5.9328 & $-$0.967 & $-$0.43 & \multicolumn{1}{c}{(b)} & 0/1 \\
12457.132 & \ion{Mg}{1} & 6.4314 & $-$2.260 & \multicolumn{1}{c}{(w)} & 0.87 & 1/1 \\
12532.840 & \ion{Cr}{1} & 2.7088 & $-$1.879 & $-$0.62 & 0.28 & 0/2 \\
12556.996 & \ion{Fe}{1} & 2.2786 & $-$3.626 & $-$0.82 & $-$0.19 & 0/2 \\
12583.924 & \ion{Si}{1} & 6.6161 & $-$0.462 & $-$0.40 & 0.12 & 0/2 \\
12589.204 & \ion{Si}{1} & 6.6161 & $-$1.507 & \multicolumn{1}{c}{(w)} & \multicolumn{1}{c}{(b)} & 0/0 \\
12600.277 & \ion{Ti}{1} & 1.4432 & $-$2.320 & $-$0.32 & 0.10 & 0/2 \\
12615.928 & \ion{Fe}{1} & 4.6382 & $-$1.517 & $-$0.83 & 0.21 & 0/2 \\
12627.674 & \ion{Si}{1} & 6.6192 & $-$0.805 & $-$0.47 & 0.09 & 0/2 \\
12638.703 & \ion{Fe}{1} & 4.5585 & $-$0.783 & $-$0.72 & 0.22 & 0/2 \\
12648.741 & \ion{Fe}{1} & 4.6070 & $-$1.140 & $-$0.63 & 0.08 & 0/2 \\
12671.096 & \ion{Ti}{1} & 1.4298 & $-$2.360 & $-$0.26 & 0.49 & 0/2 \\
12738.383 & \ion{Ti}{1} & 2.1747 & $-$1.280 & $-$0.26 & 0.27 & 0/2 \\
12743.264 & \ion{Ni}{1} & 5.2843 & $-$0.452 & $-$0.80 & $-$0.37 & 1/2 \\
12744.905 & \ion{Ti}{1} & 2.4875 & $-$1.280 & $-$0.32 & 0.01 & 0/2 \\
12789.450 & \ion{Fe}{1} & 5.0095 & $-$1.514 & $-$0.72 & 0.03 & 0/2 \\
12807.152 & \ion{Fe}{1} & 3.6398 & $-$2.452 & $-$0.61 & 0.26 & 0/2 \\
12811.478 & \ion{Ti}{1} & 2.1603 & $-$1.390 & $-$0.23 & 0.37 & 0/2 \\
12816.045 & \ion{Ca}{1} & 3.9104 & $-$0.765 & 0.17 & 0.82 & 2/2 \\
12821.672 & \ion{Ti}{1} & 1.4601 & $-$1.190 & $-$0.05 & 0.82 & 1/2 \\
12823.867 & \ion{Ca}{1} & 3.9104 & $-$0.997 & $-$0.42 & \multicolumn{1}{c}{(b)} & 0/1 \\
12824.859 & \ion{Fe}{1} & 3.0176 & $-$3.835 & $-$0.22 & \multicolumn{1}{c}{(b)} & 1/1 \\
12827.059 & \ion{Ca}{1} & 3.9104 & $-$1.478 & $-$0.33 & 0.31 & 0/2 \\
12831.445 & \ion{Ti}{1} & 1.4298 & $-$1.490 & $-$0.09 & 0.84 & 0/2 \\
12840.574 & \ion{Fe}{1} & 4.9556 & $-$1.329 & $-$0.81 & \multicolumn{1}{c}{(b)} & 0/1 \\
12847.034 & \ion{Ti}{1} & 1.4432 & $-$1.330 & $-$0.19 & 0.47 & 0/2 \\
12870.041 & \ion{Mg}{1} & 6.5879 & $-$1.530 & $-$0.33 & 0.26 & 0/2 \\
12879.766 & \ion{Fe}{1} & 2.2786 & $-$3.458 & $-$0.63 & 0.17 & 0/2 \\
12885.290 & \ion{Ca}{1} & 4.4300 & $-$1.164 & \multicolumn{1}{c}{(w)} & 0.07 & 0/1 \\
12896.118 & \ion{Fe}{1} & 4.9130 & $-$1.424 & $-$0.74 & $-$0.01 & 0/2 \\
12910.090 & \ion{Cr}{1} & 2.7079 & $-$1.779 & $-$0.58 & 0.23 & 0/2 \\
12919.899 & \ion{Ti}{1} & 2.1535 & $-$1.560 & $-$0.37 & 0.33 & 0/2 \\
12921.810 & \ion{Cr}{1} & 2.7088 & $-$2.743 & \multicolumn{1}{c}{(w)} & 0.18 & 0/1 \\
12927.477 & \ion{Ti}{1} & 2.1535 & $-$2.440 & \multicolumn{1}{c}{(w)} & $-$0.27 & 0/1 \\
12932.313 & \ion{Ni}{1} & 2.7403 & $-$2.523 & $-$0.93 & $-$0.02 & 0/2 \\
12937.020 & \ion{Cr}{1} & 2.7099 & $-$1.896 & $-$0.66 & 0.07 & 0/2 \\
12950.896 & \ion{Ti}{1} & 3.4406 & $-$0.569 & \multicolumn{1}{c}{(w)} & 0.44 & 0/1 \\
12987.567 & \ion{Ti}{1} & 2.5057 & $-$1.550 & $-$0.30 & 0.20 & 0/2 \\
13006.684 & \ion{Fe}{1} & 2.9904 & $-$3.744 & $-$0.24 & 0.93 & 2/2 \\
13011.897 & \ion{Ti}{1} & 1.4432 & $-$2.270 & $-$0.30 & 0.42 & 0/2 \\
13014.841 & \ion{Fe}{1} & 5.4457 & $-$1.693 & \multicolumn{1}{c}{(w)} & 0.46 & 0/1 \\
13033.554 & \ion{Ca}{1} & 4.4410 & $-$0.064 & $-$0.48 & 0.23 & 0/2 \\
13039.647 & \ion{Fe}{1} & 5.6547 & $-$0.731 & $-$0.96 & $-$0.47 & 0/2 \\
13048.181 & \ion{Ni}{1} & 4.5379 & $-$1.008 & $-$0.65 & 0.34 & 0/2 \\
13077.265 & \ion{Ti}{1} & 1.4601 & $-$2.220 & $-$0.24 & \multicolumn{1}{c}{(b)} & 0/1 \\
13086.027 & \ion{Si}{1} & 6.0827 & $-$1.412 & $-$0.62 & \multicolumn{1}{c}{(b)} & 0/1 \\
13098.876 & \ion{Fe}{1} & 5.0095 & $-$1.290 & $-$0.83 & $-$0.06 & 0/2 \\
13102.057 & \ion{Si}{1} & 6.0827 & $-$0.309 & $-$0.35 & $-$0.07 & 0/2 \\
13107.972 & \ion{Fe}{1} & 5.6693 & $-$1.449 & \multicolumn{1}{c}{(w)} & $-$0.37 & 0/1 \\
13134.942 & \ion{Ca}{1} & 4.4506 & 0.085 & $-$0.37 & 0.44 & 0/2 \\
13145.071 & \ion{Fe}{1} & 4.1426 & $-$3.296 & \multicolumn{1}{c}{(w)} & 0.30 & 0/1 \\
13147.920 & \ion{Fe}{1} & 5.3933 & $-$0.814 & $-$0.49 & 0.25 & 0/2 \\
13152.743 & \ion{Si}{1} & 4.9201 & $-$2.504 & $-$0.26 & 0.08 & 0/2 \\
13176.888 & \ion{Si}{1} & 5.8625 & $-$0.200 & $-$0.07 & 0.75 & 0/2 \\
13201.150 & \ion{Cr}{1} & 2.7088 & $-$1.834 & $-$0.59 & 0.25 & 0/2 \\
13212.418 & \ion{Ni}{1} & 2.7403 & $-$3.901 & \multicolumn{1}{c}{(w)} & \multicolumn{1}{c}{(b)} & 0/0 \\
13217.020 & \ion{Cr}{1} & 2.7099 & $-$2.302 & $-$0.59 & 0.65 & 0/2 \\
13222.523 & \ion{Fe}{1} & 5.6547 & $-$1.278 & \multicolumn{1}{c}{(w)} & 0.03 & 0/1 \\
13255.812 & \ion{Ti}{1} & 2.2312 & $-$2.119 & $-$0.27 & 0.13 & 0/2 \\
\enddata
\tablecomments{
Here presented is the entire table which is also available
as an online material on the journal website.
}
\end{deluxetable}

~\newpage
~\newpage

\startlongtable
\begin{deluxetable}{ccccccc}%[!tb]
\tabletypesize{\small}
\tablecaption{Lines Selected from MB99 and Abundances}
\tablehead{
  \colhead{$\lambda _{\rm air}$}
& \colhead{Atom}
& \colhead{EP}
& \colhead{$\loggf$}
& \colhead{Arcturus}
& \colhead{$\mu$~Leo}
& \colhead{flag}
\\ 
  \colhead{(\AA)}
& \colhead{}
& \colhead{(eV)}
& \colhead{(dex)}
& \colhead{(dex)}
& \colhead{(dex)}
& \colhead{}
} 
\startdata 
10003.09 & \ion{Ti}{1} & 2.16 & $-$1.32 & $-$0.22 & 0.19 & 0/2 \\
10011.74 & \ion{Ti}{1} & 2.15 & $-$1.54 & $-$0.03 & \multicolumn{1}{c}{(b)} & 0/1 \\
10013.86 & \ion{Si}{1} & 6.40 & $-$1.73 & \multicolumn{1}{c}{(w)} & $-$0.17 & 0/1 \\
10019.79 & \ion{Fe}{1} & 5.48 & $-$1.44 & \multicolumn{1}{c}{(w)} & 0.30 & 0/1 \\
10032.86 & \ion{Fe}{1} & 5.51 & $-$1.36 & \multicolumn{1}{c}{(w)} & 0.26 & 0/1 \\
10034.49 & \ion{Ti}{1} & 1.46 & $-$2.09 & 0.12 & 0.65 & 0/2 \\
10041.47 & \ion{Fe}{1} & 5.01 & $-$1.84 & \multicolumn{1}{c}{(w)} & 0.51 & 0/1 \\
10059.90 & \ion{Ti}{1} & 1.43 & $-$2.40 & 0.02 & 0.54 & 0/2 \\
10065.05 & \ion{Fe}{1} & 4.84 & $-$0.57 & $-$0.25 & 0.52 & 0/2 \\
10066.55 & \ion{Ti}{1} & 2.16 & $-$1.85 & $-$0.10 & 0.47 & 0/2 \\
10068.37 & \ion{Si}{1} & 6.10 & $-$1.40 & $-$0.24 & $-$0.12 & 0/2 \\
10080.30 & \ion{Cr}{1} & 3.56 & $-$1.45 & \multicolumn{1}{c}{(w)} & 0.30 & 0/1 \\
10081.39 & \ion{Fe}{1} & 2.42 & $-$4.53 & $-$0.43 & 0.18 & 0/2 \\
10098.55 & \ion{Si}{1} & 6.40 & $-$1.76 & \multicolumn{1}{c}{(w)} & 0.44 & 0/1 \\
10114.02 & \ion{Fe}{1} & 2.76 & $-$3.76 & $-$0.52 & \multicolumn{1}{c}{(b)} & 0/1 \\
10145.57 & \ion{Fe}{1} & 4.80 & $-$0.41 & $-$0.24 & 0.51 & 0/2 \\
10155.16 & \ion{Fe}{1} & 2.18 & $-$4.36 & $-$0.49 & 0.17 & 0/2 \\
10167.47 & \ion{Fe}{1} & 2.20 & $-$4.26 & $-$0.39 & 0.34 & 0/2 \\
10189.13 & \ion{Ti}{1} & 1.46 & $-$3.27 & \multicolumn{1}{c}{(w)} & 0.61 & 0/1 \\
10193.23 & \ion{Ni}{1} & 4.09 & $-$0.81 & $-$0.37 & 0.50 & 0/2 \\
10195.11 & \ion{Fe}{1} & 2.73 & $-$3.63 & $-$0.51 & 0.27 & 0/2 \\
10197.01 & \ion{Cr}{1} & 2.99 & $-$2.44 & \multicolumn{1}{c}{(w)} & $-$0.04 & 0/1 \\
10216.32 & \ion{Fe}{1} & 4.73 & $-$0.29 & $-$0.2: & 0.62 & 1/2 \\
10218.41 & \ion{Fe}{1} & 3.07 & $-$2.93 & $-$0.38 & 0.55 & 0/2 \\
10230.78 & \ion{Fe}{1} & 5.87 & $-$0.70 & \multicolumn{1}{c}{(w)} & 0.40 & 0/1 \\
10249.15 & \ion{Ca}{1} & 4.53 & $-$0.96 & \multicolumn{1}{c}{(w)} & 0.40 & 0/1 \\
10254.77 & \ion{Ca}{1} & 4.53 & $-$0.98 & \multicolumn{1}{c}{(w)} & 0.46 & 0/1 \\
10265.22 & \ion{Fe}{1} & 2.22 & $-$4.67 & $-$0.47 & 0.13 & 0/2 \\
10273.69 & \ion{Ca}{1} & 4.53 & $-$0.76 & \multicolumn{1}{c}{(w)} & 0.16 & 0/1 \\
10288.94 & \ion{Si}{1} & 4.92 & $-$1.71 & $-$0.31 & 0.41 & 0/2 \\
10299.24 & \ion{Mg}{1} & 6.12 & $-$2.06 & \multicolumn{1}{c}{(b)} & $-$0.83: & 1/1 \\
10301.41 & \ion{Si}{1} & 6.10 & $-$1.83 & \multicolumn{1}{c}{(w)} & 0.44 & 0/1 \\
10307.45 & \ion{Fe}{1} & 4.59 & $-$2.45 & \multicolumn{1}{c}{(w)} & 0.17 & 0/1 \\
10312.52 & \ion{Mg}{1} & 6.12 & $-$1.71 & $-$0.17 & 0.25 & 0/2 \\
10340.89 & \ion{Fe}{1} & 2.20 & $-$3.65 & $-$0.40 & 0.16 & 0/2 \\
10343.83 & \ion{Ca}{1} & 2.93 & $-$0.40 & \multicolumn{1}{c}{(s)} & \multicolumn{1}{c}{(s)} & 0/0 \\
10347.96 & \ion{Fe}{1} & 5.39 & $-$0.82 & $-$0.47 & 0.57 & 0/2 \\
10353.81 & \ion{Fe}{1} & 5.39 & $-$1.09 & \multicolumn{1}{c}{(w)} & 0.33 & 0/1 \\
10371.27 & \ion{Si}{1} & 4.93 & $-$0.80 & $-$0.06 & 0.40 & 0/2 \\
10395.80 & \ion{Fe}{1} & 2.18 & $-$3.42 & $-$0.61 & $-$0.03 & 0/2 \\
10396.81 & \ion{Ti}{1} & 0.85 & $-$1.79 & \multicolumn{1}{c}{(s)} & 0.71 & 0/1 \\
10401.72 & \ion{Fe}{1} & 3.02 & $-$4.36 & \multicolumn{1}{c}{(w)} & 0.25 & 0/1 \\
10406.96 & \ion{Si}{1} & 6.62 & $-$0.77 & $-$0.38 & $-$0.14 & 0/2 \\
10414.85 & \ion{Si}{1} & 6.62 & $-$1.38 & \multicolumn{1}{c}{(w)} & 0.25 & 0/1 \\
10416.65 & \ion{Cr}{1} & 3.01 & $-$2.40 & \multicolumn{1}{c}{(w)} & $-$0.10 & 0/1 \\
10435.36 & \ion{Fe}{1} & 4.73 & $-$2.11 & \multicolumn{1}{c}{(w)} & 0.45 & 0/1 \\
10452.75 & \ion{Fe}{1} & 3.88 & $-$2.30 & $-$0.67 & 0.27 & 0/2 \\
10469.66 & \ion{Fe}{1} & 3.88 & $-$1.37 & $-$0.43 & 0.42 & 0/2 \\
10481.27 & \ion{Ca}{1} & 4.74 & $-$0.83 & \multicolumn{1}{c}{(w)} & 0.35 & 0/1 \\
10486.22 & \ion{Cr}{1} & 3.01 & $-$1.16 & $-$0.40 & 0.46 & 0/2 \\
10496.09 & \ion{Ti}{1} & 0.84 & $-$1.91 & \multicolumn{1}{c}{(s)} & 0.49 & 0/1 \\
10509.99 & \ion{Cr}{1} & 3.01 & $-$1.78 & $-$0.47 & 0.26 & 0/2 \\
10516.14 & \ion{Ca}{1} & 4.74 & $-$0.52 & \multicolumn{1}{c}{(w)} & 0.25 & 0/1 \\
10530.52 & \ion{Ni}{1} & 4.11 & $-$1.30 & $-$0.36 & 0.58: & 1/2 \\
10532.24 & \ion{Fe}{1} & 3.93 & $-$1.76 & $-$0.35 & 0.27 & 0/2 \\
10550.06 & \ion{Cr}{1} & 3.01 & $-$2.66 & \multicolumn{1}{c}{(w)} & 0.47 & 0/1 \\
10555.65 & \ion{Fe}{1} & 5.45 & $-$1.39 & \multicolumn{1}{c}{(w)} & 0.16 & 0/1 \\
10565.97 & \ion{Ti}{1} & 2.24 & $-$2.10 & \multicolumn{1}{c}{(w)} & 0.37 & 0/1 \\
10577.14 & \ion{Fe}{1} & 3.30 & $-$3.28 & $-$0.42 & 0.55 & 0/2 \\
10582.17 & \ion{Si}{1} & 6.22 & $-$1.16 & $-$0.27 & 0.24 & 0/2 \\
10585.14 & \ion{Si}{1} & 4.95 & $-$0.06 & \multicolumn{1}{c}{(b)} & \multicolumn{1}{c}{(s)} & 0/0 \\
10603.44 & \ion{Si}{1} & 4.93 & $-$0.37 & 0.21 & \multicolumn{1}{c}{(s)} & 0/1 \\
10607.73 & \ion{Ti}{1} & 0.85 & $-$3.16 & 0.07 & 0.56 & 0/2 \\
10611.68 & \ion{Fe}{1} & 6.17 & $-$0.09 & $-$0.44 & 0.63 & 0/2 \\
10616.72 & \ion{Fe}{1} & 3.27 & $-$3.34 & $-$0.52 & 0.19 & 0/2 \\
10627.65 & \ion{Si}{1} & 5.86 & $-$0.50 & 0.09 & 0.86 & 0/2 \\
10647.65 & \ion{Cr}{1} & 3.01 & $-$1.78 & $-$0.54 & 0.24 & 0/2 \\
10660.97 & \ion{Si}{1} & 4.92 & $-$0.32 & \multicolumn{1}{c}{(s)} & \multicolumn{1}{c}{(s)} & 0/0 \\
10661.63 & \ion{Ti}{1} & 0.82 & $-$2.07 & $-$0.12 & \multicolumn{1}{c}{(b)} & 0/1 \\
10667.52 & \ion{Cr}{1} & 3.01 & $-$1.69 & $-$0.51 & 0.39 & 0/2 \\
10672.14 & \ion{Cr}{1} & 3.01 & $-$1.57 & $-$0.46 & 0.41 & 0/2 \\
10677.05 & \ion{Ti}{1} & 0.84 & $-$2.90 & 0.02 & 0.45 & 0/2 \\
10689.72 & \ion{Si}{1} & 5.95 & $-$0.09 & 0.06 & 0.73 & 0/2 \\
10694.26 & \ion{Si}{1} & 5.96 & 0.10 & $-$0.09 & 0.62 & 0/2 \\
10725.19 & \ion{Fe}{1} & 3.64 & $-$2.98 & $-$0.54 & 0.33 & 0/2 \\
10726.39 & \ion{Ti}{1} & 0.81 & $-$2.31 & 0.02 & 0.49 & 0/2 \\
10727.42 & \ion{Si}{1} & 5.98 & 0.29 & $-$0.19 & 0.42 & 0/2 \\
10732.87 & \ion{Ti}{1} & 0.83 & $-$2.82 & 0.01 & 0.54 & 0/2 \\
10749.39 & \ion{Si}{1} & 4.93 & $-$0.21 & \multicolumn{1}{c}{(s)} & \multicolumn{1}{c}{(s)} & 0/0 \\
10753.01 & \ion{Fe}{1} & 3.96 & $-$2.14 & $-$0.48 & 0.29 & 0/2 \\
10754.76 & \ion{Fe}{1} & 2.83 & $-$4.39 & $-$0.51 & 0.22 & 0/2 \\
10774.87 & \ion{Ti}{1} & 0.82 & $-$2.98 & 0.01 & 0.45 & 0/2 \\
10780.70 & \ion{Fe}{1} & 3.24 & $-$3.59 & $-$0.55 & 0.20 & 0/2 \\
10783.05 & \ion{Fe}{1} & 3.11 & $-$2.80 & $-$0.51 & $-$0.01 & 0/2 \\
10784.56 & \ion{Si}{1} & 5.96 & $-$0.72 & $-$0.33 & 0.18 & 0/2 \\
10786.87 & \ion{Si}{1} & 4.93 & $-$0.38 & \multicolumn{1}{c}{(s)} & \multicolumn{1}{c}{(s)} & 0/0 \\
10791.45 & \ion{Ca}{1} & 4.74 & $-$0.68 & \multicolumn{1}{c}{(w)} & 0.47 & 0/1 \\
10792.51 & \ion{Ti}{1} & 0.85 & $-$3.80 & $-$0.95: & $-$0.17 & 1/2 \\
10796.11 & \ion{Si}{1} & 6.18 & $-$1.49 & $-$0.20 & 0.47 & 0/2 \\
10801.36 & \ion{Cr}{1} & 3.01 & $-$1.77 & $-$0.59 & 0.38 & 0/2 \\
10809.08 & \ion{Ca}{1} & 4.68 & $-$1.08 & \multicolumn{1}{c}{(w)} & \multicolumn{1}{c}{(b)} & 0/0 \\
10816.90 & \ion{Cr}{1} & 3.01 & $-$2.01 & $-$0.40 & 0.50 & 0/2 \\
10818.28 & \ion{Fe}{1} & 3.96 & $-$2.23 & $-$0.39 & 0.56 & 0/2 \\
10821.68 & \ion{Cr}{1} & 3.01 & $-$1.73 & \multicolumn{1}{c}{(b)} & 0.26 & 0/1 \\
10827.10 & \ion{Si}{1} & 4.95 & 0.23 & \multicolumn{1}{c}{(s)} & \multicolumn{1}{c}{(s)} & 0/0 \\
10833.38 & \ion{Ca}{1} & 4.88 & $-$0.43 & \multicolumn{1}{c}{(w)} & 0.30 & 0/1 \\
10838.98 & \ion{Ca}{1} & 4.88 & 0.03 & $-$0.21 & 0.30 & 0/2 \\
10843.86 & \ion{Si}{1} & 5.86 & $-$0.05 & 0.14 & 0.73 & 0/2 \\
10846.79 & \ion{Ca}{1} & 4.74 & $-$0.64 & \multicolumn{1}{c}{(w)} & 0.48 & 0/1 \\
10849.46 & \ion{Fe}{1} & 5.54 & $-$0.73 & $-$0.46 & 0.37 & 0/2 \\
10853.00 & \ion{Fe}{1} & 3.87 & $-$3.27 & \multicolumn{1}{c}{(w)} & 0.53 & 0/1 \\
10861.59 & \ion{Ca}{1} & 4.88 & $-$0.49 & \multicolumn{1}{c}{(w)} & 0.34 & 0/1 \\
10863.52 & \ion{Fe}{1} & 4.73 & $-$1.06 & $-$0.44 & 0.35 & 0/2 \\
10869.54 & \ion{Si}{1} & 5.08 & 0.36 & \multicolumn{1}{c}{(s)} & \multicolumn{1}{c}{(s)} & 0/0 \\
10879.88 & \ion{Ca}{1} & 4.88 & $-$0.51 & \multicolumn{1}{c}{(w)} & 0.08 & 0/1 \\
10881.76 & \ion{Fe}{1} & 2.85 & $-$3.50 & $-$0.48 & 0.36 & 0/2 \\
10882.81 & \ion{Si}{1} & 5.98 & $-$0.62 & $-$0.33 & 0.11 & 0/2 \\
10884.26 & \ion{Fe}{1} & 3.93 & $-$2.18 & $-$0.51 & 0.11 & 0/2 \\
10885.35 & \ion{Si}{1} & 6.18 & $-$0.10 & $-$0.14 & 0.58 & 0/2 \\
10896.30 & \ion{Fe}{1} & 3.07 & $-$2.93 & $-$0.51 & 0.27 & 0/2 \\
10905.72 & \ion{Cr}{1} & 3.44 & $-$0.70 & $-$0.30 & 0.48 & 0/2 \\
10953.32 & \ion{Mg}{1} & 5.93 & $-$0.90 & 0.08 & 0.43 & 0/2 \\
10979.31 & \ion{Si}{1} & 4.95 & $-$0.60 & 0.29 & \multicolumn{1}{c}{(s)} & 0/1 \\
10982.08 & \ion{Si}{1} & 6.19 & $-$0.27 & $-$0.08 & 0.55 & 0/2 \\
11015.53 & \ion{Cr}{1} & 3.45 & $-$0.58 & $-$0.22: & 0.53 & 1/2 \\
11026.78 & \ion{Fe}{1} & 3.94 & $-$2.77 & $-$0.47 & 0.10 & 0/2 \\
11053.52 & \ion{Fe}{1} & 3.98 & $-$3.09 & \multicolumn{1}{c}{(w)} & $-$0.07 & 0/1 \\
11119.80 & \ion{Fe}{1} & 2.85 & $-$2.54 & $-$0.46 & \multicolumn{1}{c}{(b)} & 0/1 \\
11130.03 & \ion{Si}{1} & 6.21 & $-$0.31 & $-$0.06 & 0.14 & 0/2 \\
11135.96 & \ion{Fe}{1} & 5.31 & $-$1.10 & $-$0.31 & 0.25 & 0/2 \\
11607.57 & \ion{Fe}{1} & 2.20 & $-$2.46 & \multicolumn{1}{c}{(s)} & \multicolumn{1}{c}{(s)} & 0/0 \\
11638.26 & \ion{Fe}{1} & 2.18 & $-$2.59 & \multicolumn{1}{c}{(s)} & \multicolumn{1}{c}{(s)} & 0/0 \\
11640.96 & \ion{Si}{1} & 6.27 & $-$0.48 & $-$0.17 & 0.24 & 0/2 \\
11681.60 & \ion{Fe}{1} & 3.55 & $-$3.41 & $-$0.64 & 0.13 & 0/2 \\
11700.27 & \ion{Si}{1} & 6.27 & $-$0.67 & 0.04 & 1.45: & 1/2 \\
11715.49 & \ion{Fe}{1} & 5.64 & $-$1.20 & \multicolumn{1}{c}{(w)} & 0.35 & 0/1 \\
11780.55 & \ion{Ti}{1} & 1.44 & $-$2.42 & $-$0.06 & 0.19 & 0/2 \\
11783.26 & \ion{Fe}{1} & 2.83 & $-$1.86 & \multicolumn{1}{c}{(s)} & \multicolumn{1}{c}{(s)} & 0/0 \\
11797.18 & \ion{Ti}{1} & 1.43 & $-$2.33 & $-$0.24 & 0.22 & 0/2 \\
11828.19 & \ion{Mg}{1} & 4.35 & $-$0.50 & \multicolumn{1}{c}{(s)} & \multicolumn{1}{c}{(s)} & 0/0 \\
11863.92 & \ion{Si}{1} & 5.98 & $-$1.50 & $-$0.24 & 0.30 & 0/2 \\
11892.89 & \ion{Ti}{1} & 1.43 & $-$1.73 & $-$0.12 & 0.29 & 0/2 \\
11900.03 & \ion{Si}{1} & 5.96 & $-$1.79 & $-$0.11 & 0.48 & 0/2 \\
11927.84 & \ion{Ni}{1} & 4.27 & $-$2.26 & \multicolumn{1}{c}{(w)} & $-$0.23 & 0/1 \\
11955.95 & \ion{Ca}{1} & 4.13 & $-$0.91 & $-$0.31 & 0.30 & 0/2 \\
11973.04 & \ion{Fe}{1} & 2.18 & $-$2.28 & \multicolumn{1}{c}{(s)} & \multicolumn{1}{c}{(s)} & 0/0 \\
11984.23 & \ion{Si}{1} & 4.93 & 0.12 & \multicolumn{1}{c}{(s)} & \multicolumn{1}{c}{(s)} & 0/0 \\
11991.58 & \ion{Si}{1} & 4.92 & $-$0.22 & \multicolumn{1}{c}{(s)} & \multicolumn{1}{c}{(s)} & 0/0 \\
12000.97 & \ion{Cr}{1} & 3.44 & $-$1.93 & \multicolumn{1}{c}{(w)} & $-$0.09 & 0/1 \\
12031.53 & \ion{Si}{1} & 4.95 & 0.24 & \multicolumn{1}{c}{(s)} & \multicolumn{1}{c}{(s)} & 0/0 \\
12039.84 & \ion{Mg}{1} & 5.75 & $-$1.55 & $-$0.13 & 0.16 & 0/2 \\
12053.08 & \ion{Fe}{1} & 4.56 & $-$1.75 & $-$0.40 & 0.20 & 0/2 \\
12103.54 & \ion{Si}{1} & 4.93 & $-$0.49 & 0.28 & \multicolumn{1}{c}{(s)} & 0/1 \\
12105.84 & \ion{Ca}{1} & 4.55 & $-$0.54 & $-$0.22 & 0.34 & 0/2 \\
12119.50 & \ion{Fe}{1} & 4.59 & $-$1.88 & $-$0.46 & 0.34 & 0/2 \\
12133.99 & \ion{Si}{1} & 5.98 & $-$1.89 & $-$0.24 & 0.36 & 0/2 \\
12175.75 & \ion{Si}{1} & 6.62 & $-$0.97 & $-$0.17 & 0.49 & 0/2 \\
12178.40 & \ion{Si}{1} & 6.27 & $-$1.14 & $-$0.10 & 0.61 & 0/2 \\
12190.10 & \ion{Fe}{1} & 3.63 & $-$2.75 & $-$0.39 & 0.45 & 0/2 \\
12196.70 & \ion{Si}{1} & 5.08 & $-$3.27 & \multicolumn{1}{c}{(w)} & $-$0.19 & 0/1 \\
12213.34 & \ion{Fe}{1} & 4.64 & $-$2.09 & $-$0.30 & 0.53 & 0/2 \\
12227.12 & \ion{Fe}{1} & 4.61 & $-$1.60 & $-$0.55 & 0.17 & 0/2 \\
12255.70 & \ion{Ti}{1} & 3.92 & $-$0.07 & \multicolumn{1}{c}{(w)} & 0.20 & 0/1 \\
12270.71 & \ion{Si}{1} & 4.95 & $-$0.54 & 0.16 & \multicolumn{1}{c}{(s)} & 0/1 \\
12283.28 & \ion{Fe}{1} & 6.17 & $-$0.61 & \multicolumn{1}{c}{(w)} & 0.28 & 0/1 \\
12340.49 & \ion{Fe}{1} & 2.28 & $-$4.79 & $-$0.56 & 0.15 & 0/2 \\
12342.92 & \ion{Fe}{1} & 4.64 & $-$1.68 & $-$0.47 & 0.48 & 0/2 \\
12388.37 & \ion{Ti}{1} & 2.16 & $-$1.81 & $-$0.93: & 0.11 & 1/2 \\
12390.17 & \ion{Si}{1} & 5.08 & $-$1.93 & $-$0.24 & 0.46 & 0/2 \\
12395.84 & \ion{Si}{1} & 4.95 & $-$1.82 & $-$0.23 & 0.41 & 0/2 \\
12417.92 & \ion{Mg}{1} & 5.93 & $-$1.69 & $-$0.18 & 0.26 & 0/2 \\
12423.02 & \ion{Mg}{1} & 5.93 & $-$1.23 & $-$0.18 & 0.17 & 0/2 \\
12433.45 & \ion{Mg}{1} & 5.93 & $-$1.00 & $-$0.23 & 0.15 & 0/2 \\
12449.42 & \ion{Ni}{1} & 6.10 & $-$0.01 & \multicolumn{1}{c}{(w)} & $-$0.03 & 0/1 \\
12460.70 & \ion{Ti}{1} & 4.24 & 0.78 & $-$0.72: & 0.11 & 1/2 \\
12532.85 & \ion{Cr}{1} & 2.71 & $-$2.07 & $-$0.43 & 0.36 & 0/2 \\
12557.01 & \ion{Fe}{1} & 2.28 & $-$4.07 & $-$0.36 & 0.06 & 0/2 \\
12583.95 & \ion{Si}{1} & 6.62 & $-$0.62 & $-$0.23 & 0.30 & 0/2 \\
12589.21 & \ion{Si}{1} & 6.62 & $-$1.56 & \multicolumn{1}{c}{(w)} & 0.35 & 0/1 \\
12600.27 & \ion{Ti}{1} & 1.44 & $-$2.48 & $-$0.19 & 0.16 & 0/2 \\
12615.93 & \ion{Fe}{1} & 4.64 & $-$1.77 & $-$0.56 & 0.40 & 0/2 \\
12627.70 & \ion{Si}{1} & 6.62 & $-$1.07 & $-$0.22 & 0.33 & 0/2 \\
12638.72 & \ion{Fe}{1} & 4.56 & $-$1.00 & $-$0.39 & 0.49 & 0/2 \\
12648.74 & \ion{Fe}{1} & 4.61 & $-$1.32 & $-$0.39 & 0.20 & 0/2 \\
12671.10 & \ion{Ti}{1} & 1.43 & $-$2.19 & $-$0.40 & 0.31 & 0/2 \\
12738.39 & \ion{Ti}{1} & 2.71 & $-$0.90 & 0.10 & 0.51 & 0/2 \\
12743.26 & \ion{Ni}{1} & 5.28 & $-$0.91 & \multicolumn{1}{c}{(w)} & 0.07 & 0/1 \\
12744.91 & \ion{Ti}{1} & 2.49 & $-$1.54 & $-$0.04 & 0.32 & 0/2 \\
12789.47 & \ion{Fe}{1} & 5.01 & $-$1.92 & \multicolumn{1}{c}{(w)} & 0.39 & 0/1 \\
12807.16 & \ion{Fe}{1} & 3.64 & $-$2.76 & $-$0.29 & 0.45 & 0/2 \\
12811.48 & \ion{Ti}{1} & 2.16 & $-$1.60 & $-$0.01 & 0.50 & 0/2 \\
12816.05 & \ion{Ca}{1} & 3.91 & $-$1.27 & 0.82: & 1.19: & 2/2 \\
12821.67 & \ion{Ti}{1} & 1.46 & $-$1.67 & 0.44: & 1.02 & 1/2 \\
12824.87 & \ion{Fe}{1} & 3.02 & $-$3.68 & $-$0.47 & 0.05 & 0/2 \\
12827.02 & \ion{Ca}{1} & 3.91 & $-$1.70 & \multicolumn{1}{c}{(w)} & 0.52 & 0/1 \\
12831.41 & \ion{Ti}{1} & 1.43 & $-$1.85 & 0.29 & 0.97 & 0/2 \\
12840.58 & \ion{Fe}{1} & 4.95 & $-$1.76 & $-$0.23 & \multicolumn{1}{c}{(b)} & 0/1 \\
12847.05 & \ion{Ti}{1} & 1.44 & $-$1.71 & 0.21 & 0.65 & 0/2 \\
12879.78 & \ion{Fe}{1} & 2.28 & $-$3.61 & $-$0.44 & 0.13 & 0/2 \\
12896.12 & \ion{Fe}{1} & 4.91 & $-$1.80 & $-$0.37 & 0.30 & 0/2 \\
12910.10 & \ion{Cr}{1} & 2.71 & $-$1.99 & $-$0.46 & 0.27 & 0/2 \\
12919.90 & \ion{Ti}{1} & 2.15 & $-$1.74 & $-$0.19 & 0.45 & 0/2 \\
12921.81 & \ion{Cr}{1} & 2.71 & $-$2.73 & \multicolumn{1}{c}{(w)} & 0.20 & 0/1 \\
12937.03 & \ion{Cr}{1} & 2.71 & $-$2.09 & $-$0.48 & 0.15 & 0/2 \\
12946.54 & \ion{Fe}{1} & 3.25 & $-$4.23 & \multicolumn{1}{c}{(w)} & $-$0.01 & 0/1 \\
12950.90 & \ion{Ti}{1} & 3.44 & $-$0.54 & \multicolumn{1}{c}{(w)} & 0.40 & 0/1 \\
13001.42 & \ion{Ca}{1} & 4.44 & $-$1.24 & \multicolumn{1}{c}{(w)} & $-$0.08 & 0/1 \\
13006.70 & \ion{Fe}{1} & 2.99 & $-$3.49 & $-$0.53 & 0.42 & 0/2 \\
13011.90 & \ion{Ti}{1} & 1.44 & $-$2.50 & $-$0.10 & 0.55 & 0/2 \\
13014.85 & \ion{Fe}{1} & 5.45 & $-$1.68 & \multicolumn{1}{c}{(w)} & 0.30 & 0/1 \\
13033.56 & \ion{Ca}{1} & 4.44 & $-$0.31 & $-$0.23 & 0.36 & 0/2 \\
13039.66 & \ion{Fe}{1} & 5.66 & $-$1.32 & \multicolumn{1}{c}{(w)} & 0.09 & 0/1 \\
13077.27 & \ion{Ti}{1} & 1.46 & $-$2.34 & $-$0.10 & \multicolumn{1}{c}{(b)} & 0/1 \\
13086.44 & \ion{Ca}{1} & 4.44 & $-$0.90 & $-$0.36 & 0.25 & 0/2 \\
13098.92 & \ion{Fe}{1} & 5.01 & $-$1.73 & $-$0.38 & 0.37 & 0/2 \\
13102.07 & \ion{Si}{1} & 6.08 & $-$0.72 & 0.06 & 0.30 & 0/2 \\
13134.94 & \ion{Ca}{1} & 4.45 & $-$0.14 & $-$0.16 & 0.46 & 0/2 \\
13147.93 & \ion{Fe}{1} & 5.39 & $-$0.93 & $-$0.34 & 0.34 & 0/2 \\
13152.74 & \ion{Si}{1} & 4.92 & $-$2.58 & $-$0.22 & 0.04 & 0/2 \\
13167.78 & \ion{Ca}{1} & 4.45 & $-$1.23 & \multicolumn{1}{c}{(w)} & 0.19 & 0/1 \\
13176.91 & \ion{Si}{1} & 5.86 & $-$0.30 & 0.09 & 0.80 & 0/2 \\
13201.16 & \ion{Cr}{1} & 2.71 & $-$2.08 & $-$0.34 & 0.37 & 0/2 \\
13212.45 & \ion{Ni}{1} & 2.74 & $-$3.04 & $-$0.10 & 0.8: & 1/2 \\
13217.02 & \ion{Cr}{1} & 2.71 & $-$2.52 & $-$0.38 & 0.73 & 0/2 \\
13264.17 & \ion{Ni}{1} & 5.28 & $-$1.03 & \multicolumn{1}{c}{(w)} & $-$0.27 & 0/1 \\
13291.78 & \ion{Fe}{1} & 5.48 & $-$1.58 & \multicolumn{1}{c}{(w)} & 0.25 & 0/1 \\
\enddata
\tablecomments{
Here presented is the entire table which is also available
as an online material on the journal website.
}
\end{deluxetable}

\end{document}